\documentclass[11pt,a4paper]{article}
\pdfoutput=1 

\usepackage{jheppub}

\usepackage[normalem]{ulem}
\usepackage[latin1]{inputenc}
\usepackage{graphicx}
\graphicspath{{Figures/}}
\usepackage{latexsym,epsfig,amssymb, amsmath,nicefrac}
\usepackage{xcolor}
\usepackage{subcaption}
\usepackage{slashed}
\usepackage{tikz}
\usepackage{comment}
\usepackage{color}
\usepackage[makeroom]{cancel}
\usepackage{latexsym}
\usepackage{epsf}
\usepackage{amssymb}
\usepackage{graphicx}
\usepackage{amsmath}
\usepackage{amsmath,amssymb,amsthm}
\usepackage{verbatim}
\usepackage{hyperref}
\usepackage{physics}
\usepackage{float}

\newcommand{\e}{e}

\def\tb0{\tilde{\beta}_0}
\def\b0{\beta_0}

\def\bi{\begin{itemize}}
\def\ei{\end{itemize}}
\def\be{\begin{equation}}
\def\ee{\end{equation}}
\newcommand{\bea}{\begin{eqnarray}}
\newcommand{\eea}{\end{eqnarray}}

\def\mpl{M_{\rm Pl}}
\def\dd{{\rm d}}
\def\pd{\partial}
\def\calP{{\cal P}}
\def\calR{{\cal R}}
\def\calS{{\cal S}}
\def\prk{\calP_\calR}

\def\calN{{\cal N}}
\def\calG{{\cal G}}

\def\calM{{\cal M}}
\def\calD{{\cal D}}

\renewcommand{\b}[1]{\left(#1\right)}

\title{Spectator Axions in String Inflation and Primordial Black Holes}

\author[a,b]{Michele Cicoli,}
\author[c,d]{Dario L. Lorenzoni,}
\author[d]{Evan McDonough,}
\author[a,b]{Francisco G. Pedro}

\affiliation[a]{\small Dipartimento di Fisica e Astronomia, Universit\`a di Bologna, via Irnerio 46, 40126 Bologna, Italy}
\affiliation[b]{\small INFN, Sezione di Bologna, viale Berti Pichat 6/2, 40127 Bologna, Italy}
\affiliation[c]{\small Department of Physics \& Astronomy, University of Manitoba, Winnipeg, MB R3T 2N2, Canada}
\affiliation[d]{\small Department of Physics, University of Winnipeg, Winnipeg, MB R3B 2E9, Canada}

\emailAdd{michele.cicoli@unibo.it}
\emailAdd{lorenzod@myumanitoba.ca}
\emailAdd{e.mcdonough@uwinnipeg.ca}
\emailAdd{francisco.soares@unibo.it}

\abstract{We study the impact of light spectator axions on the seeding of primordial black holes (PBHs) during inflation in string theory. Primordial black holes exhibit unique and novel phenomenology, and may constitute the observed dark matter. Cosmic inflation provides a mechanism for producing them, but such inflation models typically feature Planckian field excursions, necessitating an ultraviolet completion into quantum gravity. String theory provides a natural framework for doing so, and indeed Fibre Inflation has been shown to produce PBHs while satisfying constraints from cosmic microwave background data. In this work we study the dynamics of axions during Fibre Inflation, and find a diverse and rich set of possibilities, including turns in field space and enhancement of primordial perturbations. We find that across most of parameter space, notably an axion with a far sub-Planckian decay constant $f\ll \mpl$, there is a negligible impact on the power spectrum of curvature perturbations, indicating an overall robustness of the model. On the other hand, an axion with a larger but still sub-Planckian decay constant, $f\gtrsim {\cal O}(0.1) \mpl$, and an exponentially small prefactor of its non-perturbative potential, can enhance the growth of perturbations, making it easier to achieve the amplification needed to seed PBHs, effectively realizing {\it axion-assisted} PBHs in string theory.}

\begin{document}

\maketitle

\section{Introduction}
\label{sec:intro}

Primordial black holes (PBHs) \cite{Zeldovich:1967lct,Hawking:1971ei,Carr:1974nx,Meszaros:1974tb,Carr:1975qj,Khlopov:1985jw,Niemeyer:1999ak}  
have emerged as a prominent candidate to explain the observed dark matter abundance \cite{Khlopov:2008qy,Carr:2009jm,Sasaki:2018dmp,Carr:2020gox,Carr:2020xqk,Green:2020jor,Escriva:2021aeh,Villanueva-Domingo:2021spv,Escriva:2022duf,Gorton:2024cdm}. 
PBHs in the so-called asteroid mass range, with $10^{17} \, {\rm g} \leq M_{\rm PBH} \leq 10^{23} \, {\rm g}$, could account for the entirety of the dark matter in the Universe. PBHs may play a role in explaining a host of other observations, such as high-redshift supermassive black holes \cite{Volonteri:2021sfo,Dayal:2024zwq,Hai-LongHuang:2024vvz,Hai-LongHuang:2024gtx,Huang:2024aog}, gravitational waves at LIGO \cite{Bird:2016dcv}, NANOGrav \cite{DeLuca:2020agl,DeLorenci:2025wbn}, and future detectors \cite{Qin:2023lgo}, as well as ultrahigh-energy neutrinos \cite{Klipfel:2025jql,Baker:2025cff} and other cosmic rays \cite{Klipfel:2025bvh}, and more. The open question in all the studies of PBHs is their origin: What seeded the formation of PBHs during the first moments of the Universe?

A natural possibility is that the origin of PBHs lies in the same physics thought to be responsible for the formation of structure in the Universe: quantum fluctuations during cosmic inflation. Among many different proposals to realize PBH dark matter via cosmic inflation, a well-studied class of models posits a transient phase of ``ultra-slow roll'' (USR) inflation \cite{Kinney:2005vj,Martin:2012pe,Ezquiaga:2017fvi,Garcia-Bellido:2017mdw,Germani:2017bcs,Kannike:2017bxn,Motohashi:2017kbs,Di:2017ndc,Ballesteros:2017fsr,Pattison:2017mbe,Passaglia:2018ixg,Biagetti:2018pjj,Mishra:2019pzq,Figueroa:2020jkf,Karam:2022nym,Ozsoy:2023ryl,Cole:2023wyx,Cicoli:2018asa,Cicoli:2022sih,Cai:2022erk,Inomata:2021uqj,Inomata:2021tpx,Bhaumik:2019tvl,Pi:2022ysn,Choudhury:2024one}, wherein a rapid deceleration of the inflaton triggers a growth of the perturbations that exit the Hubble radius around that time. In order for models to match constraints from cosmic microwave background (CMB) data on large scales while also producing asteroid-mass PBHs, these models typically feature a Planckian field excursion. Indeed, the decoupling of CMB and PBH scales is achieved by a large distance in field space between different regions of the inflationary potential. Large field inflation poses a significant challenge to field theory model building, since operators allowed by effective field theory, even if Planck suppressed, can spoil the flatness of the potential when $\Delta \phi \gtrsim \mpl$. This manifests what is known as the $\eta$ problem (for a textbook discussion see, e.g., \cite{Baumann:2014nda}).

String theory provides a UV complete framework to describe cosmic inflation, allowing for overall consistency tests and for explicit calculations of leading and subleading corrections to the inflaton potential (see \cite{Cicoli:2023opf} for a comprehensive review on string cosmology). Conversely, cosmic inflation provides a testing ground for string theory, with constraints from the CMB reframed as constraints on the geometry and topology of the extra dimensions. Inflationary model building further provides an opportunity to identify idiosyncratic features of string theory, e.g., in the form of a potential energy functional which appears {\it ad hoc} in the context of field theory but emerges naturally from a string compactification. For all these reasons, inflation remains a highly active area of string theory research.

Primordial black holes seeded by inflation provide another window into testing models of inflation based in string theory. A promising proposal to do so is provided by Refs.~\cite{Cicoli:2018asa,Cicoli:2022sih} in the context of Fibre Inflation \cite{Cicoli:2008gp,Broy:2015zba,Cicoli:2016chb,Cicoli:2016xae, Cicoli:2017axo, Burgess:2016owb,Cicoli:2018cgu,Cicoli:2020bao,Cicoli:2022uqa,Cicoli:2024bxw}, which involves a type IIB compactification on a fibred Calabi-Yau manifold where the inflaton is identified with the ratio between two K\"ahler moduli. Refs.~\cite{Cicoli:2018asa,Cicoli:2022sih} established that the general form of the scalar potential in Fibre Inflation easily admits, for suitable parameter values, a near-inflection point. This can cause a phase of ultra-slow roll inflation and allow for the production of PBHs.

A characteristic aspect of Fibre Inflation which was not analyzed in detail in Refs.~\cite{Cicoli:2018asa,Cicoli:2022sih} is the presence of two axions which are lighter than the inflation and behave as spectator fields during inflation. This is a manifestation of the generic prediction of string theory of a spectrum of axion-like particles, the so-called string {\it axiverse} \cite{Arvanitaki:2009fg,Cicoli:2012sz}. 
The search for axions as a test of string theory is its own burgeoning area of research, with applications to several phenomena from black hole superradiance \cite{Mehta:2021pwf} to detection at axion helioscopes \cite{Halverson:2019cmy} and more. 

In the context of inflation, axion-like particles are expected to be merely {\it spectator fields}, namely a subdominant contribution to the energy density driving inflation, unimportant to the dynamics of the background evolution and of the adiabatic curvature perturbations\footnote{Exceptions to this are natural inflation \cite{Freese:1990rb} and related models, wherein an axion-like particle is identified with the inflaton.
}. However, spectator fields can play an important role in PBH models featuring ultra-slow roll, as shown in Refs.~\cite{Lorenzoni:2025gni,Lorenzoni:2025kwn}. As explained in those works, spectator fields can have a dramatic impact on PBH models. In fact, even a very small contribution to the total energy density, and hence to the Hubble parameter, can increase the Hubble friction felt by the inflaton enough to significantly enhance the production of PBHs. This relaxes the need to fine-tune the feature in the inflaton potential to ensure PBH production, leading to an overall {\it resilience} of USR inflation PBH models to parameter variations.

In this work we examine the production of PBHs from Fibre Inflation when including the two light axions. The string compactification fixes the masses, decay constants, and couplings of the axions. Unlike the field theory models of Refs.~\cite{Lorenzoni:2025gni,Lorenzoni:2025kwn}, the spectator axions are intrinsically coupled to the inflaton, both through their kinetic terms and through the potential. This leads to rich dynamics in the combined inflaton-axions system that are absent in the decoupled case. 

We find that the results of Refs.~\cite{Cicoli:2018asa,Cicoli:2022sih} are robust to the inclusion of axions. Throughout most of parameter space, the axions leave the power spectrum of curvature perturbations unchanged. For certain tuned parameter choices, we find instead results similar to those reported in Refs.~\cite{Lorenzoni:2025gni,Lorenzoni:2025kwn} since the axions can dramatically enhance the peak of the curvature power spectrum. This axion-induced enhancement relaxes the parameter fine-tuning needed to engineer the near-inflection point in the inflaton potential. These results reaffirm the status of Fibre Inflation as a promising framework for realizing PBHs from cosmic inflation.

This work builds on and extends Refs.~\cite{Cicoli:2018ccr,Cicoli:2019ulk,Cicoli:2021itv,Cicoli:2021yhb}, which studied isocurvature perturbations from spectator axions in Fibre Inflation. Those works demonstrated that the axions of Fibre Inflation easily satisfy CMB isocurvature constraints. The novelty of the present work lies in its focus on PBHs. We study a class of Fibre Inflation models featuring a transient phase of ultra-slow roll, and analyze the evolution of perturbations with particular attention to the small scales relevant to PBHs. 

The structure of this paper is as follows. In Sec.~\ref{sec:fibre} we derive the 4D effective field theory for Fibre Inflation including the axions. In Sec.~\ref{sec:pbh}, we review the production of PBHs from the single-field realization of the model as presented in Refs.~\cite{Cicoli:2018asa,Cicoli:2022sih}. We then bring the axions into the model. In Sec.~\ref{sec:axions} we study the background evolution, and find a rich interplay of the two sectors, with the axions slowing down the inflaton and the inflaton causing a dramatic swing in the axion masses and their resulting evolution. In Sec.~\ref{sec:pert} we examine the dynamics of curvature and isocurvature perturbations in this model, followed by Sec.~\ref{sec:power-spectrum} where we present results for the power spectrum including an example where the axions induce a significant enhancement of the peak of the power spectrum, realizing axion-assisted PBH production from inflation in string theory. We close in Sec.~\ref{sec:disc} with a discussion of directions for future work.

\section{Effective Field Theory of Fibre Inflation and its Axions}
\label{sec:fibre}

Fibre inflation \cite{Cicoli:2008gp,Broy:2015zba,Cicoli:2016chb,Cicoli:2016xae, Cicoli:2017axo, Burgess:2016owb,Cicoli:2018cgu,Cicoli:2020bao,Cicoli:2022uqa,Cicoli:2024bxw} is a prominent proposal to describe the first moments of the Universe in the context of string theory. Here we review the construction of the model, with particular attention to the axion fields which are usually assumed to be non-dynamical during inflation.

We consider a type IIB compactification with 3 K\"ahler moduli, $T_i=\tau_i+i\theta_i$, for $i=1,2,3$, where $\tau_i$ are $4$-cycle volumes and $\theta_i$ are $C_4$ axions, $\theta^i = \int_{\Sigma^i_4} C_4$, where $C_4$ is the Ramond-Ramond $4$-form and $\Sigma_4 ^i$  is a basis of $4$-cycles.  We consider a Calabi-Yau manifold with a fibred structure, with volume given by: 
\begin{equation}
    \mathcal{V}=\sqrt{\tau_1}\tau_2-\tau_3^{3/2}\,,    
\end{equation}
where $\tau_1$ is the volume of a 4D K3 fibration over a 2D $\mathbb{P}^1$ base with volume $\tau_2/\sqrt{\tau_1}$, while $\tau_3$ is the volume of a small cycle, or blow-up mode. This type of compactification has been extensively studied in the context of inflation \cite{Cicoli:2008gp,Broy:2015zba,Cicoli:2016chb,Cicoli:2016xae, Cicoli:2017axo, Burgess:2016owb,Cicoli:2018cgu,Cicoli:2020bao,Cicoli:2022uqa,Cicoli:2024bxw}, dark energy \cite{Cicoli:2012tz,Cicoli:2021skd,Cicoli:2024yqh}, early dark energy \cite{Cicoli:2023qri}, and dark matter model building \cite{Cicoli:2021gss}.

The 4D effective field theory is described by a K\"ahler potential $K$, which takes the general form
\begin{equation}
    K= K_{\rm K\ddot{a}hler}+K_{\rm cs}+K_{\rm dilaton}\,,
\end{equation}
where the three contributions correspond to the K\"ahler moduli, the complex structure moduli, and the axio-dilaton. The latter two are stabilized by fluxes, and in what follows we focus on the K\"ahler moduli. The K\"ahler potential then reads
\begin{equation}
    K = -2\ln\left(\mathcal{V}+\frac{\xi}{2 g_s^{3/2}}\right) + K_{g_s}\,,
\end{equation}
where $\xi$ is an $\mathcal{O}(1)$ constant controlling $\mathcal{O}(\alpha'^3)$ corrections \cite{Becker:2002nn} and the string coupling $g_s$ is assumed to be fixed by fluxes in the perturbative regime where $g_s\lesssim 0.1$. $K_{g_s}$ represents string loop corrections \cite{vonGersdorff:2005bf,Berg:2005ja,Berg:2007wt,Cicoli:2007xp,Gao:2022uop}. We omit the `uplift' of the global vacuum from AdS to dS -- which can be accomplished by a variety of mechanisms -- since the main focus of the present work is on the dynamics of inflation. Finally, there are additional higher-derivative $F^4$ corrections \cite{Ciupke:2015msa} which are however non-K\"ahler, i.e., they can be written down directly as corrections to the scalar potential, but not to $K$. 

The K\"ahler potential fixes the kinetic sector of the theory. The Lagrangian for the kinetic terms is given by:
\begin{equation}
\label{eq:Lkin}
    \mathcal{L}_{\rm kin} = \frac{1}{4\tau_1^2}\left[\left(\partial\tau_1\right)^2+\left(\partial\theta_1\right)^2\right] +\frac{1}{2\tau_2^2}\left[\left(\partial\tau_2\right)^2+\left(\partial\theta_2\right)^2\right] +\frac{3}{8\mathcal{V}\sqrt{\tau_3}}\left[\left(\partial\tau_3\right)^2+\left(\partial\theta_3\right)^2\right]\,.
\end{equation}
The scalar potential for the theory is generated by the superpotential $W$, which reads:
\begin{equation}
    W=W_0 + \sum_{i=1}^3 A_i\,e^{-a_i T_i}\,, 
\end{equation}
where the constant $W_0$ corresponds to the Gukov-Vafa-Witten flux induced superpotential \cite{Gukov:1999ya},  
which we take to be ${\cal O}(1)$ and negative, i.e. $W_0=-|W_0|$, and the second term in $W$ corresponds to non-perturbative corrections, which we assume to arise from stringy instantons or gaugino condensation on D7 branes wrapping 4-cycles. The $A_i$'s are all expected to be $\mathcal{O}(1)$ constants (in Planck units), while $a_i=2\pi/N_i$ with $N_i\in \mathbb{N}$. 

The resulting scalar potential is given by the usual expression \cite{Freedman:2012zz}
\begin{equation}\label{eqn:SUGRApotential}
    V = e^{K}\b{D_{I} W K^{I\bar{J}} D_{\bar{J}} \overline{W}-3|W|^2} \,,
\end{equation}
where $I$ labels the chiral superfields, $D_I$ is the K\"ahler covariant derivative, and $K^{I\bar{J}}$ is the inverse of the K\"ahler metric $K_{I\bar{J}}\equiv \partial_I \partial_{\bar{J}}K$. 
This is in general a function of all the fields, $\{\tau_i,\theta_i\}$ for $i=1,2,3$. However, in the limit of a large volume, $\mathcal{V}\gg 1$, and with a hierarchy $\tau_1,\tau_2 \gg \tau_3$, the scalar potential admits a perturbative expansion with a leading term which depends only on $\mathcal{V}$, $\tau_3$ and $\theta_3$, a subleading term which encodes the dependence on $u\equiv \tau_1/\tau_2$, and sub-subleading contributions which encode the dependence on the axions $\theta_1$ and $\theta_2$. Concretely,
\begin{equation}\label{eq:V-fibre-terms}
    V= V_{\rm lead}(\mathcal{V},\tau_3,\theta_3) + V_{\rm sub}(u) + V_{\rm sub-sub}(\theta_1,\theta_2)\,,    
\end{equation}
where we have written down explicitly the dependence of each contribution on the fields which are stabilised by that correction.

The leading term $V_{\rm lead}(\mathcal{V},\tau_3,\theta_3)$ is the standard LVS potential generated by the correction to $K$ proportional to $\xi$ and the contribution to $W$ proportional to $A_3$ \cite{Balasubramanian:2005zx,Conlon:2005ki,Cicoli:2008va}. At this level of approximation, the overall volume is fixed to be exponentially large as $\mathcal{V}\sim W_0\,e^{a_3\tau_3}$ with $\tau_3\sim g_s^{-1}$. The axion $\theta_3$ is stabilised such that $\cos(a_3\theta_3)=1$. Consequently, $\tau_3$ and $\theta_3$ are the heaviest moduli, with a mass of order the gravitino mass: 
$m_{\tau_3}\sim m_{\theta_3}\sim m_{3/2} \sim |W_0|/\mathcal{V}$, in Planck units ($\mpl=1$). The mass of the volume mode is more $\mathcal{V}$-suppressed since it is given by $m_{\mathcal{V}}\sim |W_0|/\mathcal{V}^{3/2}$, again in Planck units.

The subleading correction $V_{\rm sub}(u)$ lifts the remaining flat direction $u\equiv \tau_1/\tau_2$ in the saxionic moduli space, and takes the same form as Eq. (2.8) of Ref.~\cite{Cicoli:2018asa} with $\tau_{\rm K3}\equiv \tau_1 = \left(\mathcal{V}\,u\right)^{2/3}$. The mass squared of $u$ around the minimum is $m_u^2\sim |W_0|^2 /\left(\mathcal{V}^3\sqrt{\tau_1}\right)$. Finally, the sub-subleading potential $V_{\rm sub-sub}(\theta_1,\theta_2)$ lifts the two axions $\theta_1$ and $\theta_2$.

Given that the kinetic Lagrangian, Eq.~\eqref{eq:Lkin}, is not canonical, $\mathcal{V}$ and $u$ are not the canonically normalized fields. Ignoring the heavy fields $\tau_3$ and $\theta_3$ which decouple, the mass eigenstates with canonical kinetic terms are $\sigma$ and $\phi$. These relate to the original moduli as:
\begin{equation}
    \tau_1= e^{\sqrt{\frac23}\sigma+\frac{2}{\sqrt{3}}\phi} \, , \qquad
    \tau_2= e^{\sqrt{\frac23}\sigma-\frac{1}{\sqrt{3}}\phi} \,,
\end{equation}
or
\begin{equation}
    \mathcal{V}= e^{\sqrt{\frac32}\sigma}\, ,\qquad u = \frac{\tau_1}{\tau_2}= e^{\sqrt{3}\phi} \,.
\end{equation}

In the context of Fibre Inflation, $\phi$ is identified with the inflaton, while $\sigma$ is a heavy spectator field during inflation. The axion fields $\theta_1$ and $\theta_2$ are light spectator fields. The inflationary potential takes the same form as Eq. (2.9) of \cite{Cicoli:2018asa}, which corresponds to Eq. (\ref{eq:V-inf}) in Sec.~\ref{sec:pbh}, where the inflaton has been shifted from its minimum as
\begin{equation}\label{eq:phi-shift}
    \phi=\langle\phi\rangle+\varphi \,,
\end{equation} 
which implies:
\begin{equation}
    \tau_1 = \mathcal{V}^{2/3} e^{\frac{2}{\sqrt{3}}\phi}\equiv\langle\tau_1\rangle\,e^{\frac{2}{\sqrt{3}}\varphi}\,, \qquad
    \tau_2 = \mathcal{V}^{2/3} e^{-\frac{1}{\sqrt{3}}\phi} \equiv \langle\tau_2\rangle\,e^{-\frac{1}{\sqrt{3}}\varphi} \,.
\end{equation}
During inflation, $\mathcal{V}$ remains constant around $\mathcal{V}\sim\mathcal{O}(10^{3-4})$ in order to match the observed amplitude of the density perturbations, while $\phi$ evolves from larger to smaller values. Hence $\tau_1$ evolves from larger to smaller values, while $\tau_2$ evolves from smaller to larger values in such a way that $\mathcal{V}$ is kept constant. The typical inflaton field range in Fibre Inflation is $\Delta\var  phi\sim\mathcal{O}(5)\,\mpl$. This can be easily rewritten in terms of the canonically unnormalized moduli as
\begin{equation}
    \Delta\varphi = \frac{\sqrt{3}}{2}\ln\left(\frac{\tau_{1,*}}{\tau_{1,{\rm end}}}\right)\qquad\Leftrightarrow\qquad \tau_{1,*} = \tau_{1,{\rm end}}\,e^{\frac{2}{\sqrt{3}}\Delta\varphi} \,,
\end{equation}
where $\tau_{1,*}$ denotes the value of $\tau_1$ at CMB horizon exit, around $N_e\simeq 52$ e-foldings before the end of inflation \cite{Cicoli:2018cgu}, while $\tau_{1,{\rm end}}$ denotes the value of $\tau_1$ when inflation ends, i.e. when $\varepsilon\simeq 1$.

The two axions $\theta_1$ and $\theta_2$ can be written in terms of the corresponding canonically normalized axions $\chi_1$ and $\chi_2$ around the minimum as
\begin{equation}\label{eq:chi-def}
    \chi_1=\frac{2\pi}{N_1}\,f_1\,\theta_1\,,\qquad \qquad    
    \chi_2=\frac{2\pi}{N_2}\,f_2\,\theta_2\,,
\end{equation}
where the two axion decay constants are given by
\begin{equation}
\label{eq:fa-def}
    f_1=\frac{N_1}{2\sqrt{2}\pi\langle\tau_1\rangle}\,,\qquad\qquad
    f_2=\frac{N_2}{2\pi\langle\tau_2\rangle}    \,.
\end{equation}
Hence the axionic kinetic Lagrangian becomes
\begin{equation}
    {\cal L}_{\rm kin} \supset \frac{1}{2}e^{-\frac{4}{\sqrt{3}}\varphi} (\partial \chi_1)^2 + \frac{1}{2}e^{+\frac{2}{\sqrt{3}}\varphi} (\partial \chi_2)^2 \,.
\end{equation}
The axion potential is given by $V_{\rm sub-sub}$ and takes the form
\begin{equation}\label{eq:V-sub-sub}
    V_{\rm sub-sub}(\varphi,\theta_1,\theta_2) = \Lambda_1(\varphi) \left[1-\cos\left(\frac{\chi_1}{f_1}\right)\right]+ \Lambda_2(\varphi)\left[1-\cos\left(\frac{\chi_2}{f_2}\right)\right] \,,
\end{equation}
where $\Lambda_{1,2}(\varphi)$ are inflaton-dependent:
\begin{align}
    \Lambda_1(\varphi) &\equiv \frac{2\sqrt{2} |W_0| A_1}{\mathcal{V}^2 f_1}\,\,e^{\frac{2}{\sqrt{3}}\varphi}\,e^{-\frac{1}{\sqrt{2}f_1}\,e^{\frac{2}{\sqrt{3}}\varphi}} \,,
    \label{eq:Lambda-1}\\
    \Lambda_2(\varphi) &\equiv \frac{4 |W_0| A_2}{\mathcal{V}^2 f_2}\,\,e^{-\frac{1}{\sqrt{3}}\varphi}\,e^{-\frac{1}{f_2} e^{-\frac{1}{\sqrt{3}}\varphi}} \,.
    \label{eq:Lambda-2}
\end{align}
These two crucially differ by the signs in the exponential, which will play an important role in the dynamics of the two axions during inflation.

Putting these pieces together, we arrive at the low-energy effective field theory for Fibre Inflation including the axions. This is given by
\begin{equation}
\label{action-full}
    S= \int \dd^4 x \sqrt{ -g} \left[ \frac{M_\text{Pl}^2}{2} R
    - \frac{1}{2} (\partial \varphi)^2 
    - \frac{1}{2} \e^{-\frac{4}{\sqrt{3}}\varphi}(\partial \chi_1)^2 
    - \frac{1}{2} \e^{+\frac{2}{\sqrt{3}}\varphi}(\partial \chi_2)^2
    - V(\varphi,\chi_1,\chi_2)
    \right]\,,
\end{equation}
where $\varphi$ is the inflaton as identified above, while $\chi_1$ and $\chi_2$ are the canonically normalized axion fields, as defined in Eq.~\eqref{eq:chi-def}. The field potential is given by the subleading and sub-subleading terms of Eq.~\eqref{eq:V-fibre-terms}, 
\begin{equation}
\label{eq:V-tot}
    V(\varphi,\chi_1,\chi_2) = V_{\rm inf}(\varphi) + V_{\rm ax}(\chi_1,\chi_2,\varphi)\,.
\end{equation}
The inflaton potential $V_{\rm inf}$ is the conventional Fibre Inflation potential given by Eq. (2.9) of \cite{Cicoli:2018asa} and Eq. (\ref{eq:V-inf}) in Sec.~\ref{sec:pbh}, while the potential for the axions is that of Eq.~\eqref{eq:V-sub-sub}.

\section{Primordial Black Holes from Fibre Inflation}
\label{sec:pbh}

We now turn our attention to the dynamics of Fibre Inflation. We restrict our analysis to the single field limit, where the axions are assumed to have vanishing initial values and henceforth remain non-dynamical.
The inflationary potential can be written schematically as (see e.g.,~Refs.~\cite{Cicoli:2018asa,Cicoli:2022sih})
\begin{equation}
\label{eq:V-inf}
    V_{\rm inf}(\varphi) = V_0 \left[ C_1 - e^{-\frac{1}{\sqrt{3}} \varphi} \left( 1 - \frac{C_2}{1 - C_3\, e^{-\frac{1}{\sqrt{3}} \varphi}} \right) 
    + e^{\frac{2}{\sqrt{3}} \varphi} \left( C_4 - \frac{C_5}{1 + C_6\, e^{\sqrt{3}\,\varphi}} \right) \right],
\end{equation}
where $V_0\simeq m_u^2$ and $C_i$, $i=1,...,6$ are flux-dependent tunable parameters. Moreover, the inflaton is shifted as in Eq.~\eqref{eq:phi-shift}, such that $\varphi$ is zero at the minimum of its potential.
Importantly, this model can satisfy CMB constraints on the primordial power spectrum on large scales while generating a sufficiently large enhancement of perturbations on small scales to seed primordial black holes.

\begin{figure}[ht!]
    \centering
    \includegraphics[width=\textwidth]{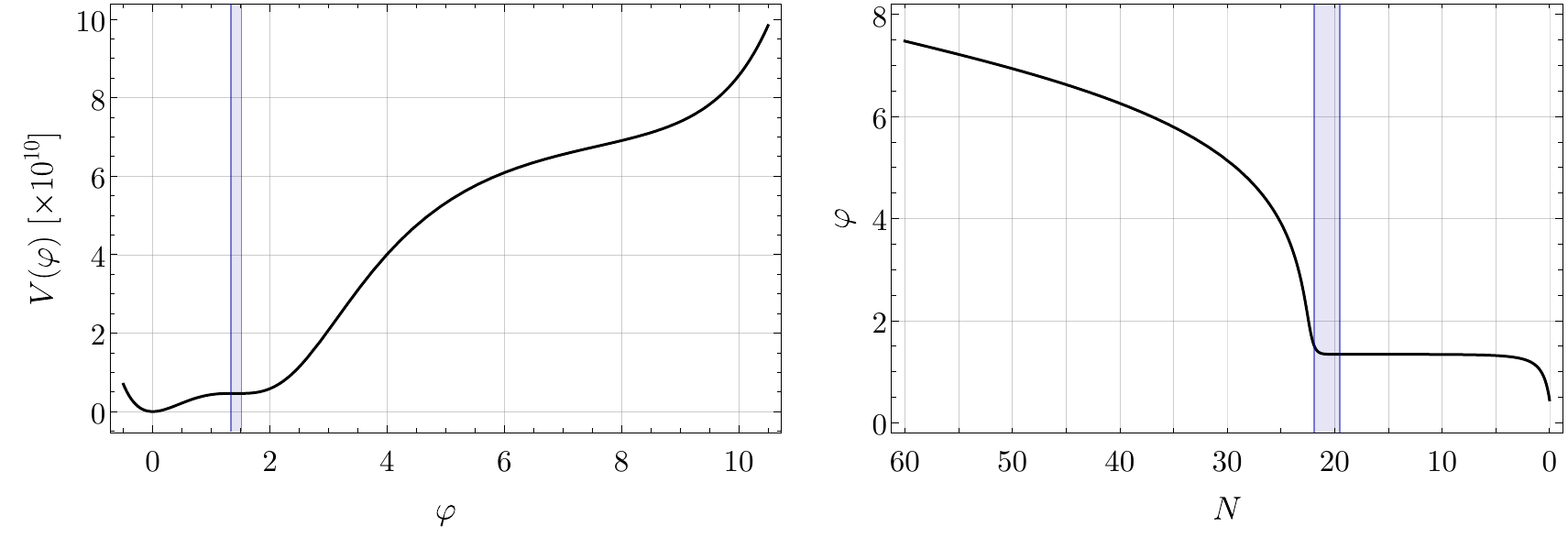}
    \caption{
    Inflationary potential and evolution of the inflaton for the Fibre Inflation model of Eq.~\eqref{eq:V-inf}. Both $V$ and $\varphi$ are in Planck units. We consider the parameter set ${\cal P}_1$ of Tab.~\ref{tab:SF-params}. 
    \textit{Right panel:} The plateau of $V$ around $\varphi \sim 1.5 \,\mpl$, highlighted in blue, is responsible for the phase of ultra-slow roll inflation and the consequent production of PBHs.
    \textit{Left panel:} The time evolution of $\varphi$ is shown in terms of the number of e-folds before the end of inflation; $\varphi$ also plateaus in the USR region.
    }
    \label{fig:V-fibre}
\end{figure}

\begin{table}[ht!]
    \centering
    \begin{tabular}{@{\extracolsep{0pt}}c|*{6}c}
        \textbf{Model}& $V_0$ & $C_2$ & $C_3$ & $C_4$ & $C_5$ & $C_6$ \\ \hline
        \textbf{Ref. }$\boldsymbol{(\calP_1)}$ & $2.21\times10^{-9}$  & $0.5$ & $0.2661248$ & $7.82\times10^{-7}$ & $0.03911638$ & $0.03575049$
        \\
        \textbf{No-PBH }$\boldsymbol{(\calP_2)}$ & $1.89\times10^{-9}$  & $0.5$ & $0.2661400$ & $7.00\times10^{-7}$ & $0.03911680$ & $0.03575230$
    \end{tabular}
    \caption{Parameter values for the single-field models considered in this work, for the inflationary potential \eqref{eq:V-inf}. The parameter $C_1$ is found by requiring $V_{\rm inf}=0$ at $\varphi=0$, such that $C_1 = 1 - C_2/(1-C_3) - C_4 + C_5/(1+C_6)$.
    The ``Reference'' parameter values $\calP_1$ reproduce the $\calP_1$ model of Ref.~\cite{Cicoli:2022sih}, and are considered throughout Sec.~\ref{sec:pbh}. The ``No-PBH'' model $\calP_2$ does not produce PBHs, and will be used in later sections. The initial conditions for the dynamics of these models are $\varphi_i=10\,\mpl$ for $\calP_1$, and $\varphi_i=8\,\mpl$ for $\calP_2$.
    }
    \label{tab:SF-params}
\end{table}

\begin{figure}
    \centering
    \includegraphics[width=\textwidth]{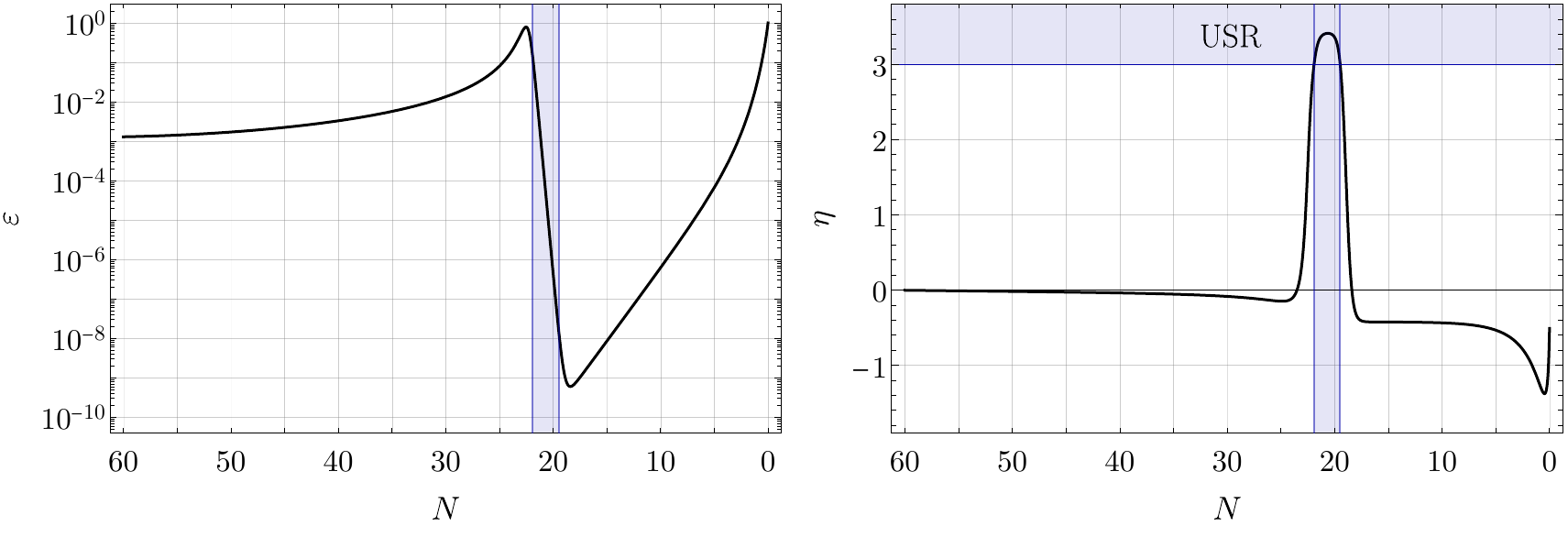}
    \caption{
    Slow-roll parameters $\varepsilon$ and $\eta$ for the Fibre Inflation model of Eq.~\eqref{eq:V-inf}, in Planck units. We consider the parameter set ${\cal P}_1$ of Tab.~\ref{tab:SF-params}. 
    \textit{Left panel:} $\varepsilon$ becomes tiny during the USR phase (in blue).
    \textit{Right panel:} The USR region (in blue) is defined by $\eta>3$.
    }
    \label{fig:SR-fibre}
\end{figure}

In Fig.~\ref{fig:V-fibre} we show a fiducial example of the inflaton potential, with parameters given in Tab.~\ref{tab:SF-params}. The plateau around $\varphi \sim 1.5\, \mpl$, highlighted in blue, is responsible for a transient phase of USR inflation. USR has been extensively studied as a mechanism to seed primordial black holes  \cite{Kinney:2005vj,Martin:2012pe,Ezquiaga:2017fvi,Garcia-Bellido:2017mdw,Germani:2017bcs,Kannike:2017bxn,Motohashi:2017kbs,Di:2017ndc,Ballesteros:2017fsr,Pattison:2017mbe,Passaglia:2018ixg,Biagetti:2018pjj,Mishra:2019pzq,Figueroa:2020jkf,Karam:2022nym,Ozsoy:2023ryl,Cole:2023wyx,Cicoli:2018asa,Cicoli:2022sih,Cai:2022erk,Inomata:2021uqj,Inomata:2021tpx,Bhaumik:2019tvl,Pi:2022ysn,Choudhury:2024one}. 
The background evolution during inflation can be characterized by the Hubble slow-roll parameters,
\begin{equation}
    \varepsilon \equiv - \frac{\dot{H}}{H^2} \,  , \qquad  \eta \equiv  2 \varepsilon - \frac{\dot{\varepsilon}}{2 \varepsilon H}  \,.
\end{equation}
USR inflation corresponds to the case $\varepsilon\ll1$ and $\eta > 3$, in contrast with conventional slow-roll inflation where both $\varepsilon, \eta \ll 1$.
The evolution of $\varepsilon$ and $\eta$ in Fibre Inflation, with parameters given in Tab.~\ref{tab:SF-params}, is shown in Fig.~\ref{fig:SR-fibre}. One may appreciate that $\varepsilon$ first increases during inflation, before undergoing a sharp decrease of roughly 8 orders of magnitude. The drop in $\varepsilon$ corresponds to a sharp spike in $\eta$, inducing a period of ultra-slow roll inflation.

The CMB observables of this model, such as $n_s$ and $r$, must be computed numerically. For the $\calP_1$ parameters in Tab.~\ref{tab:SF-params}, we find the CMB observables reported in Tab.~\ref{tab:P1-obs}, where we fix $N_{\rm CMB}=53$ following Refs.~\cite{Cicoli:2018cgu,Cicoli:2022sih}: notably, $n_s= 0.965$ and $r=0.02$. This is consistent with CMB data from the combination of Planck, SPT, and ACT \cite{Balkenhol:2025wms} \footnote{We note that the model is flexible enough to accommodate DESI in the joint dataset, which slightly raises $n_s$ for reasons explained in Refs. \cite{McDonough:2025lzo,Ferreira:2025lrd}. We leave a detailed exploration of this to future work.}.

\begin{table*}
    \centering
    \makebox[\textwidth][c]{
    \begin{tabular}{@{\extracolsep{0pt}}c|*{7}c|*{1}c}
        \textbf{Model}
        & $A_s$ & $n_s$ & $\alpha_s$ & $r$ & $\beta_{\rm iso}$ & $f_{\rm NL}^{\rm equil}$ & $M_{\rm PBH}~
        [{\rm g}]$
        & $\prk^{\rm peak}$ 
        \\ 
        \hline 
        \textbf{Ref. }$\boldsymbol{(\calP_1)}$ 
        & $2.10\times 10^{-9}$ & $0.9651$ & $-0.003$ & $0.02$ & $-$ & $109$ & $1.2\times 10^{22}$
        & $7.2\times 10^{-3}$ 
    \end{tabular}
    }
    \caption{Observables for the parameter set ${\cal P}_1$ as defined in Tab.~\ref{tab:SF-params}. CMB scales exit at $N_{\rm CMB} = 53$ \cite{Cicoli:2022sih}. Observables are computed numerically using \texttt{PyTransport}~\cite{Mulryne:2016mzv,Ronayne:2017qzn}.
    }
    \label{tab:P1-obs}
\end{table*}

This model can also seed the formation of primordial black holes. PBHs are seeded by an amplification of cosmological perturbations on small scales, manifesting as a spike in the primordial power spectrum of the comoving curvature perturbation ${\cal R}$. For a review of perturbation theory and its application to PBHs, we refer the reader to the reviews \cite{Byrnes:2021jka,Ozsoy:2023ryl}. In this work we take $\calP_{\cal R}(k_{\rm PBH}) > 10^{-3}$ as a threshold for PBH formation due to the large perturbation with comoving wave number $k_{\rm PBH}$, where $\calP_{\cal R}(k)$ is the dimensionless power spectrum, related to the Fourier modes of the comoving curvature perturbation, $\calR_k$, by
\begin{equation}\label{eq:PRk-def}
    \calP_{\cal R}(k,N) \equiv \frac{k^3}{2\pi^2}|{\cal R}_k(N)|^2\,.
\end{equation}
We numerically solve for the evolution of ${\cal R}_k(N)$, which at linear order in perturbation theory satisfy
\begin{equation}\label{eq:EoM-R-SF}
    \ddot{\calR}_k + (3+\delta)H\dot{\calR}_k + \frac{k^2}{a^2}\calR_k = 0\, ,
\end{equation}
where $\delta\equiv \dd\log \varepsilon /\dd N = 4\varepsilon-2\eta$. We evaluate this at the end of inflation ($N=0$) to construct the primordial power spectrum $\calP_{\cal R}(k)$.

\begin{figure}[ht!]
    \centering
    \includegraphics[width=0.6\textwidth]{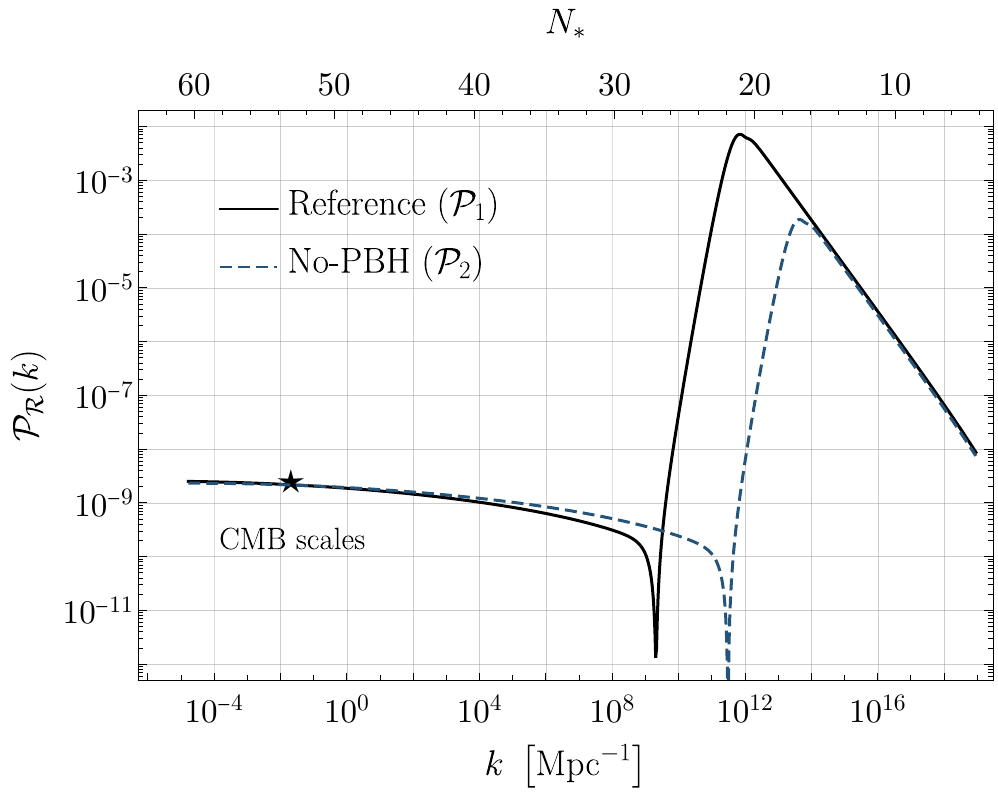}
    \caption{
    Curvature power spectrum for the Fibre Inflation model of Eq.~\eqref{eq:V-inf}, for the parameter set ${\cal P}_1$ of Tab.~\ref{tab:SF-params}. The peak of $\prk(k)$ surpasses the threshold for the production of PBHs on asteroid-mass scales. Also shown for reference is the curvature power spectrum for the $\calP_2$ model of Tab.~\ref{tab:SF-params}: In this case, $\prk(k)$ is not sufficiently enhanced to allow for the production of PBHs.
    }
    \label{fig:PRk-SF}
\end{figure}

The power spectrum for Fibre Inflation with the $\calP_1$ parameters of Tab.~\ref{tab:SF-params} is shown in Fig.~\ref{fig:PRk-SF}. One may appreciate a spike in the power spectrum to ${\cal P}_{\cal R}(k_{\rm PBH}) \approx 7  \times 10^{-3}$, well above the threshold for PBH production. The resultant PBHs have a mass of $10^{22}$ g\footnote{We determine the mass at the time of formation, $M_{\rm PBH}$, from the wavenumber $k_{\rm PBH}$ as 
\begin{equation}
    \frac{M_{\rm PBH}}{30\, M_{\odot}} \approx \left(\frac{\gamma}{0.2}\right)\left(\frac{g_{*}\left(T_{\rm f}\right)}{106.75}\right)^{-\frac{1}{6}}\left(\frac{k_{\rm PBH}}{3.2 \times 10^{5}\, \mathrm{Mpc}^{-1}}\right)^{-2}\,.
\end{equation}
We take $\gamma=0.2$ \cite{Carr:1975qj} and $g_*=106.75$ at the time of formation.
}, which is within the asteroid mass range, $10^{17} \, {\rm g} \leq M_{\rm PBH} \leq 10^{23} \, {\rm g}$ \cite{Khlopov:2008qy,Carr:2009jm,Sasaki:2018dmp,Carr:2020gox,Carr:2020xqk,Green:2020jor,Escriva:2021aeh,Villanueva-Domingo:2021spv,Escriva:2022duf,Gorton:2024cdm}, 
and are therefore a viable candidate to explain the entirety of the observed dark matter.

Finally, this model is testable by complementary gravitational wave signals in the cosmic microwave background and direct detection experiments such as LISA, ET or BBO \cite{Cicoli:2022sih}. The imprint on the CMB is parametrized by the tensor-to-scalar ratio $r=0.02$, which is within reach of the Simons Observatory's target sensitivity of $r=1.2\times 10^{-3}$ \cite{SimonsObservatory:2025avm}, while the gravitational waves sourced by the peak of the power spectrum can be probed by LISA, ET or BBO. These signals are largely insensitive to whether the PBHs constitute the observed dark matter, and remain detectable even if the PBH contribution to dark matter is exponentially suppressed \cite{Cicoli:2022sih}.

\section{Enter the Axions: Background Evolution}
\label{sec:axions}

We now revisit the dynamics of Fibre Inflation, allowing for the dynamical evolution of the axion fields during inflation. The model has been described in Sec. \ref{sec:fibre}. We restate here the most relevant equations for ease of reading. The action given in Eq. (\ref{action-full}) looks like 
\begin{equation}
\label{eq:action-full}
    S= \int \dd^4 x \sqrt{ -g} \left[ \frac{M_\text{Pl}^2}{2} R
    - \frac{1}{2} (\partial \varphi)^2 
    - \frac{1}{2} \e^{-\frac{4}{\sqrt{3}}\varphi}(\partial \chi_1)^2 
    - \frac{1}{2} \e^{+\frac{2}{\sqrt{3}}\varphi}(\partial \chi_2)^2
    - V(\varphi,\chi_1,\chi_2)
    \right]\,,
\end{equation}
where $\chi_1$ and $\chi_2$ are the canonically normalized axions around the minimum (at $\varphi=0$), as defined in Eq.~\eqref{eq:chi-def}. The scalar potential is given by Eq. (\ref{eq:V-tot}) and reads
\begin{equation}
    V(\varphi,\chi_1,\chi_2) = V_{\rm inf}(\varphi) + V_{\rm ax}(\chi_1,\chi_2,\varphi)\,.
\end{equation}
The inflaton potential $V_{\rm inf}$ is given in Eq. (\ref{eq:V-inf}), 
\begin{equation}
\label{eq:V-inf-new}
    V_{\rm inf}(\varphi) = V_0 \left[ C_1 - e^{-\frac{1}{\sqrt{3}} \varphi} \left( 1 - \frac{C_2}{1 - C_3\, e^{-\frac{1}{\sqrt{3}} \varphi}} \right) 
    + e^{\frac{2}{\sqrt{3}} \varphi} \left( C_4 - \frac{C_5}{1 + C_6\, e^{\sqrt{3}\,\varphi}} \right) \right]\,,
\end{equation}
while the potential for the two light axions is that of Eq.~\eqref{eq:V-sub-sub},
\begin{equation}\label{eq:V-ax}
    V_{\rm ax}(\chi_1,\chi_2,\varphi) = \Lambda_1(\varphi) \left[1-\cos\left(\frac{\chi_1}{f_1}\right)\right]+ \Lambda_2(\varphi)\left[1-\cos\left(\frac{\chi_2}{f_2}\right)\right] \,,
\end{equation}
with
\begin{align}
    \Lambda_1(\varphi) &\equiv \frac{2\sqrt{2} |W_0| A_1}{\mathcal{V}^2 f_1}\,\,e^{\frac{2}{\sqrt{3}}\varphi}\,e^{-\frac{1}{\sqrt{2}f_1}\,e^{\frac{2}{\sqrt{3}}\varphi}} \,,
    \label{eq:Lambda-1-new}\\
    \Lambda_2(\varphi) &\equiv \frac{4 |W_0| A_2}{\mathcal{V}^2 f_2}\,\,e^{-\frac{1}{\sqrt{3}}\varphi}\,e^{-\frac{1}{f_2} e^{-\frac{1}{\sqrt{3}}\varphi}} \,.
    \label{eq:Lambda-2-new}
\end{align}

The background evolution is determined by the Friedmann equation for the Hubble parameter and the Euler-Lagrange equations for the fields. The former is given by
\begin{equation}
\label{eq:Friedmann}
    H^2 = \frac{1}{3\mpl^2}\,\rho \,,
\end{equation}
where the energy density $\rho$ is a sum of the contributions of the inflaton and the axions,
\begin{equation}
    \rho  = \rho_{\rm inf} + \rho_{\rm ax} \,,
\end{equation}
with
\begin{eqnarray}
    \rho_{\rm inf} && \equiv  \frac{1}{2}\dot\varphi^2+V_{\rm inf}\,,\\
    \rho_{\rm ax} && \equiv  \frac{1}{2}\e^{-\frac{4}{\sqrt{3}}\varphi}\dot\chi_1^2 + \frac{1}{2}\e^{+\frac{2}{\sqrt{3}}\varphi}\dot\chi_2^2 + V_{\rm ax}\,.
\end{eqnarray}
We restrict our attention to the case where the axions make a subdominant contribution to the total energy density of the Universe, $\rho_{\rm ax} \ll \rho_{\rm inf}$, and therefore ostensibly `spectate' the dynamics of inflation driven by $\varphi$. However, this spectator status does not relegate the axions to irrelevance. This can be appreciated from the slow-roll parameter $\varepsilon$, which is given by the sum of the field contributions
\begin{equation}\label{eq:epsilon-full}
    \varepsilon  \equiv-\frac{\dot H}{H^2}
    = \varepsilon_\varphi + \varepsilon_{\chi_1} + \varepsilon_{\chi_2}\,,
\end{equation}
where
\begin{equation}
    \varepsilon_\varphi\equiv \frac{1}{2}\frac{\dot\varphi^2}{\mpl^2 H^2} 
    \,,\quad
    \varepsilon_{\chi_1} \equiv \frac{1}{2}\e^{-\frac{4}{\sqrt{3}}\varphi}\frac{\dot\chi_1^2}{\mpl^2 H^2}
    \,,\quad
    \varepsilon_{\chi_2} \equiv  \frac{1}{2}\e^{+\frac{2}{\sqrt{3}}\varphi}\frac{\dot\chi_2^2}{\mpl^2 H^2}\,.
\end{equation} 
Clearly $\varepsilon \geq \varepsilon_{\chi_1},\varepsilon_{\chi_2} $ and hence the axions effectively set a {\it floor} for the Hubble slow-roll parameter. 

\begin{figure}[ht!]
    \centering
    \includegraphics[width=\linewidth]{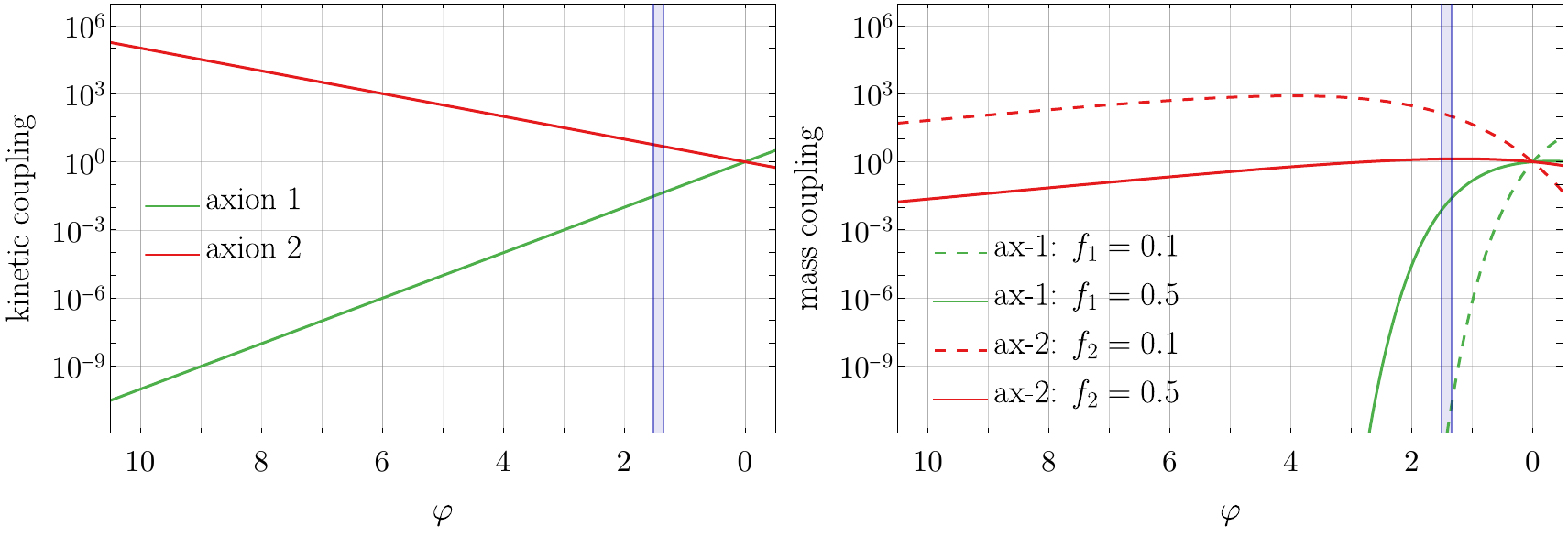}
    \caption{
    Comparison between the $\varphi$-dependence of the kinetic coupling of the axions (\textit{left panel}) and their mass coupling (\textit{right panel}), for $\chi_1$ (in green) and $\chi_2$ (in red). As the inflaton value decreases, the kinetic coupling of $\chi_1$ increases while the one of $\chi_2$ decreases. The behavior of the mass coupling depends on the value of the decay constant: for $\chi_1$, it always increases (more or less quickly) as $\varphi$ decays; for $\chi_2$, the coupling can be either increasing or decreasing at the time of the USR feature (highlighted in blue).
    }
    \label{fig:axion-phi-dependence}
\end{figure}

The equations of motion for the fields are given by
\begin{align}
    \ddot\varphi & + 3H\dot\varphi = -\partial_\varphi V_{\rm eff}\,,\label{eq:EoM-phi}\\
    \ddot\chi_1 & + \left( 3H-\frac{4}{\sqrt{3}}\dot\varphi \right)\dot\chi_1 = \e^{+\frac{4}{\sqrt{3}}\varphi}\partial_{\chi_1}V_{\rm ax}\,,\label{eq:EoM-chi1}\\
    \ddot\chi_2 & + \left( 3H+\frac{2}{\sqrt{3}}\dot\varphi \right)\dot\chi_2 = \e^{-\frac{2}{\sqrt{3}}\varphi}\partial_{\chi_2}V_{\rm ax}\,,\label{eq:EoM-chi2}
\end{align}
where we defined an effective scalar potential for the inflaton, given by
\begin{equation}
    V_{\rm eff}(\varphi,\chi_1,\chi_2)=V_{\rm inf}(\varphi) + V_{\rm ax}(\varphi,\chi_1,\chi_2) - \frac{1}{2}\left(\e^{-\frac{4}{\sqrt{3}}\varphi} \dot\chi_1^2 + \e^{+\frac{2}{\sqrt{3}}\varphi}\dot\chi_2^2 \right)\,.
\end{equation}
At this point we can identify several important effects in the combined inflaton-axion dynamics. We begin with the impact of the axions:
\begin{itemize}
    \item \textbf{Extra Hubble friction:} The energy density in the axions contributes to the Friedmann equation, Eq.~\eqref{eq:Friedmann}, which impacts the evolution of the inflaton through the Hubble friction term $H \dot{\varphi}$. This increased Hubble friction acts to slow down the inflaton beyond the usual deceleration due to the plateau feature of the potential.
    
    \item \textbf{$\boldsymbol{\varepsilon}$ floor:} Since $\varepsilon \geq \varepsilon_{\chi_1},\varepsilon_{\chi_2}$, the axions can effectively set a {\it floor} for the Hubble slow-roll parameter $\varepsilon$ whenever $\varepsilon_{\chi_1},\varepsilon_{\chi_2}\geq\varepsilon_\varphi$. 

    \item \textbf{Effective inflaton potential:} The axions contribute to the inflaton potential both through their kinetic coupling and through the potential coupling, $\Lambda(\varphi)$ in Eq.~\eqref{eq:V-ax}. These modify $V_{,\,\varphi}$ and thereby lift $\varepsilon_{\varphi}\sim (V' _{\rm eff}/V_{\rm eff})^2$. From the kinetic contributions we find the lower bound $\varepsilon_{\varphi} \gtrsim \varepsilon_\chi ^2$, while from the potential contributions we find $\varepsilon_\varphi \gtrsim (\rho_{\rm ax}/\rho_{\rm inf})^2$. 
   \end{itemize}
Conversely, the inflaton has important impacts on the axions:
\begin{itemize}
    \item \textbf{Varying axion mass:} The axion masses are inflaton-dependent and are given by $m^2 = \Lambda(\varphi)/f^2$ where $\Lambda(\varphi)$ is given in Eqs.~\eqref{eq:Lambda-1-new}-\eqref{eq:Lambda-2-new}. Concretely,
    \begin{equation}
    \label{eq:axion-mass}
        m_1^2(\varphi)  = \frac{2\sqrt{2} |W_0| A_1}{\mathcal{V}^2 f_1 ^3}\,\,e^{\frac{2}{\sqrt{3}}\varphi}\,e^{-\frac{1}{\sqrt{2}f_1}\,e^{\frac{2}{\sqrt{3}}\varphi}} \quad
        m_2^2(\varphi) = \frac{4 |W_0| A_2}{\mathcal{V}^2 f_2^3}\,\,e^{-\frac{1}{\sqrt{3}}\varphi}\,e^{-\frac{1}{f_2} e^{-\frac{1}{\sqrt{3}}\varphi}}\,.
    \end{equation}
    The axion masses around the minimum of the inflaton potential can be simply obtained by setting $\varphi=0$:
    \begin{equation}
    \label{eq:axion-mass-0}
        M_1^2 = \frac{2\sqrt{2} |W_0| A_1}{\mathcal{V}^2 f_1^3}\,e^{-\frac{1}{\sqrt{2}f_1}} \,,\qquad 
        M_2^2 = \frac{4 |W_0| A_2}{\mathcal{V}^2 f_2^3}\,e^{-\frac{1}{f_2}}\,.
    \end{equation}
    The exponential sensitivity to the decay constant indicates that an axion with a sub-Planckian decay constant is naturally light.
    
    \item \textbf{Varying kinetic couplings:} The kinetic coupling in Eq.~\eqref{eq:action-full} exhibits a different sign for the two axions, being given by $\e^{-\frac{4}{\sqrt{3}}\varphi}$ and $\e^{+\frac{2}{\sqrt{3}}\varphi}$ for $\chi_1$ and $\chi_2$, respectively. The kinetic coupling defines an instantaneous effective decay constant of the axion as $f_{{\rm eff},i} = e^{(b_i/2) \varphi}f_i$, $i=1,2$, where $b_1=-4/\sqrt{3}$ and $b_2=2/\sqrt{3}$ for $\chi_1$ and $\chi_2$, respectively. The effective decay constant is therefore increasing vs. decreasing during inflation for $\chi_1$ vs. $\chi_2$.
\end{itemize}
These two effects are illustrated in Fig.~\ref{fig:axion-phi-dependence}. It can be seen that, while the kinetic couplings of $\chi_1$ and $\chi_2$ have opposite dependence on $\varphi$ at all times, the mass couplings of both axions, if $f_2$ for $\chi_2$ is large enough, can increase as the value of the inflaton decreases. Indeed, the behavior of the axion mass depends sensitively on the decay constant, as this determines whether the exponential vs. double-exponential term dominates in Eq.~\eqref{eq:axion-mass}. For $\chi_2$, the value of $f_2$ determines whether the mass coupling will be increasing (dashed-red curve in the right panel of Fig.~\ref{fig:axion-phi-dependence}) or decreasing (solid-red) at the time of the ultra-slow roll evolution (highlighted in blue).

\begin{figure}[ht!]
    \centering
    \includegraphics[width=\linewidth]{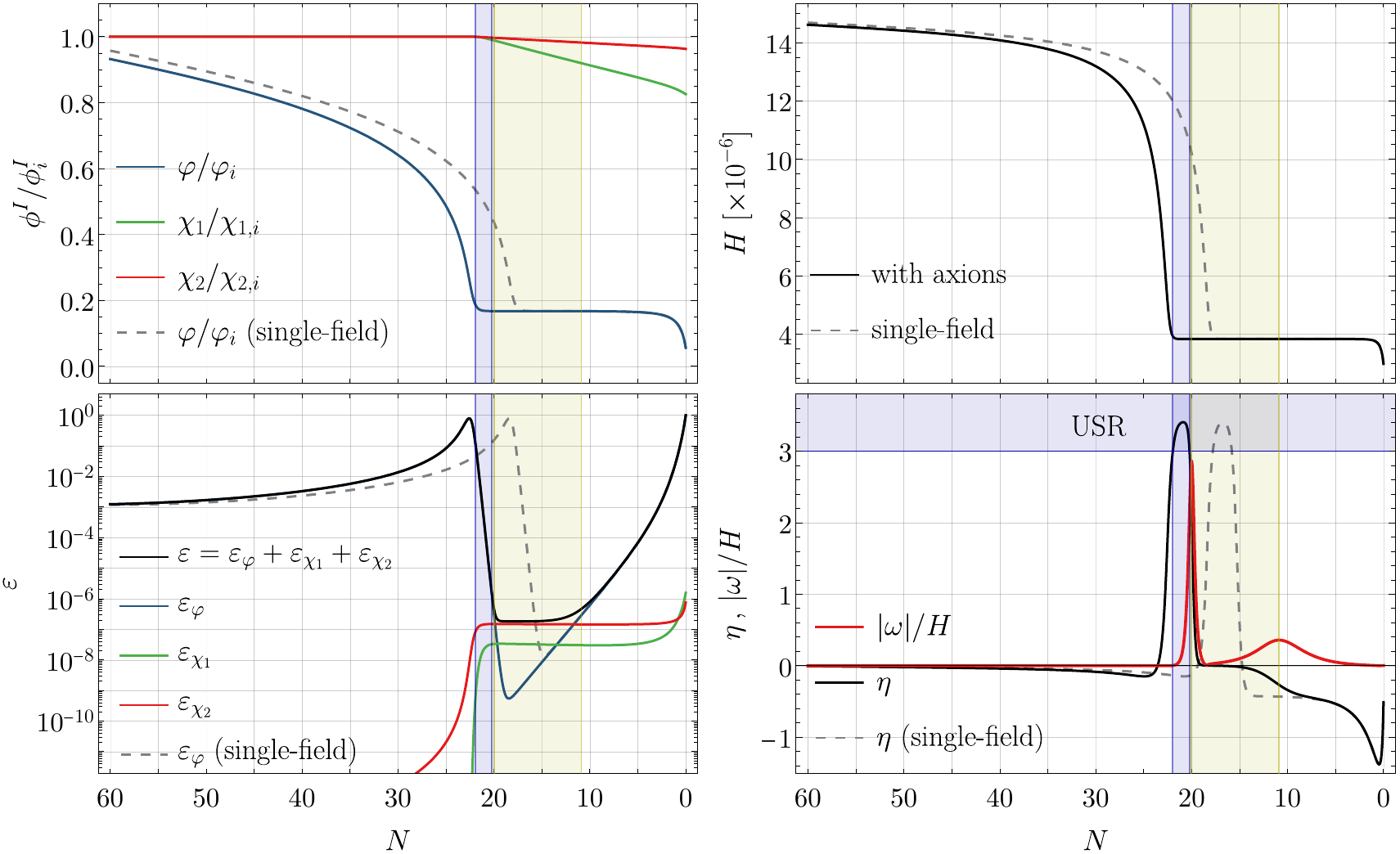}
    \caption{
    Evolution of the background quantities in the inflaton-axions system for the Fibre Inflation model of Eq.~\eqref{eq:action-full}, with the $\calP_2$ inflaton parameters given in Tab.~\ref{tab:SF-params} and the ``Illustrative'' axion parameters given in Tab.~\ref{tab:ILL-params}.
    These parameter values are taken as an illustrative example to highlight the impact of the axion fields on the inflationary dynamics.
    The blue highlight indicates the USR phase, defined by $\eta > 3$; the yellow highlight indicates the ``$\varepsilon$-floor'' phase, defined by $\varepsilon_\chi > \varepsilon_\varphi$ and delimited by the two turns in field-space.
    }
    \label{fig:ILL-bckgr}
\end{figure}

With these physical effects in mind, we solve the background equations of motion numerically and present the results in Fig.~\ref{fig:ILL-bckgr}. 
Following the conventions of the multifield inflation literature \cite{Sasaki:1995aw,Gordon:2000hv,Wands:2002bn,Langlois:2008mn,Peterson:2010np,Gong:2011uw,Kaiser:2012ak,Gong:2016qmq}, with $\phi^I=\{\varphi,\chi_1,\chi_2\}$ for $I=1,2,3$, we define the tangent vector to the field space trajectory,
\begin{equation}\label{eq:sigma_hat}
    \hat{\sigma}^I \equiv \frac{\dot{\phi}^I}{\dot{\sigma}} \,,
\end{equation}
and the magnitude of the field space velocity vector,
\begin{equation}
    \dot{\sigma} \equiv \vert \dot{\phi}^I \vert = \sqrt{ {\cal G}_{IJ} \dot{\phi}^I \dot{\phi}^J}\,,
\end{equation}
where ${\cal G}_{IJ} $ is the field-space metric. The rate of change of $\hat{\sigma}$ defines a turn rate vector,
\begin{equation}
    \omega^I\equiv {\cal D}_t\hat{\sigma}^I\,,
\end{equation}
where ${\cal D}_t \equiv \dot{\phi}^J {\cal D}_{J}$ is a directional covariant derivative in field space\footnote{The usual covariant derivative acting on a field-space vector is
    \begin{equation}\label{eq:covariant-derivative}
        \calD_J A^I\equiv \pd_J A^I + \Gamma^I_{\;JK} A^K\,,
    \end{equation}
where $\Gamma^I_{\;JK}(\phi^L)$ are the field-space Christoffel symbols and $\pd_J A^I\equiv\pd A^I/\pd\phi^J$.
}.

The top left panel of Fig.~\ref{fig:ILL-bckgr} shows the evolution of the fields, both in the single-field case and in the case where the axions are included. Comparing the evolution of the inflaton in the two models, it is clear that $\varphi$ lingers at the feature of the potential for a longer duration when the axions are present. This is a simple consequence of the additional Hubble friction, which acts to slow down the inflaton. This is mirrored in the evolution of $\varepsilon$, shown in the bottom left panel of Fig.~\ref{fig:ILL-bckgr}. Comparing $\varepsilon$ in the single-field case to $\varepsilon_\varphi$ in the multifield case, one may appreciate that $\varepsilon_\varphi$ becomes smaller and stays smaller longer when the axions are included.

\begin{table}[ht!]
    \centering
    \begin{tabular}{@{\extracolsep{5pt}}c|*{6}c}
       \textbf{Model} & $M_1$ & $f_1$ & $\chi_{1,i}$ & $M_2$ & $f_2$ & $\chi_{2,i}$ \\ \hline
       \textbf{Axion-1 ($\boldsymbol{\chi_1}$)} & $8\times10^{-7}$  & $0.5$ & $0.1\pi f_1$ & $-$ & $-$ & $-$ \\
       \textbf{Axion-2 ($\boldsymbol{\chi_2}$)} & $-$  & $-$ & $-$ & $5\times10^{-7}$ & $0.5$ & $0.1 \pi f_2$ \\
       \textbf{Illustrative} & $8\times10^{-7}$  & $0.5$ & $0.1\pi f_1$ & $5\times10^{-7}$ & $0.5$ & $0.1 \pi f_2$ \\
    \end{tabular}
    \caption{Parameter values for the axionic potential~\eqref{eq:V-ax}, for the illustrative axion models. The parameters $M_{1,2}$ represent the axion masses around the minimum of $V_{\rm inf}$, as defined in Eq.~\eqref{eq:axion-mass-0}.
    All these models use the $\calP_2$ parameters of Tab.~\ref{tab:SF-params} for the base inflationary potential~\eqref{eq:V-inf-new}.
    }
    \label{tab:ILL-params}
\end{table}

The axions themselves are near-constant at all times during inflation. This is immediately apparent from the top left panel of Fig.~\ref{fig:ILL-bckgr}. However, closer inspection, e.g., of $\varepsilon$ reveals a more complicated story: the axion slow-roll parameters $\varepsilon_{\chi}$ are negligible at early times but undergo a dramatic spike as $\varphi$ passes the feature at $\varphi\sim \mpl$. This is the combined effect of the mass and kinetic couplings, both of which influence $\varepsilon_{\chi}$. This rapid increase is such that $\varepsilon_\chi$ surpasses $\varepsilon_\varphi$ for a transient period, setting a floor on the total value of $\varepsilon$ and causing the trajectory of the field system to {\it turn} and temporarily move along the $\chi$-direction. These turns are manifest as spikes in the turn rate, as one can see from the bottom right panel of Fig.~\ref{fig:ILL-bckgr}, where we plot the magnitude of the turn-rate vector. 

The net result of these effects is a three-phase structure to the background evolution, delimited by the crossing points of $\varepsilon_{\chi}$ and $\varepsilon_{\varphi}$: at early and late times (first and third phases), $\varepsilon_{\varphi} \gg \varepsilon_{\chi}$, indicating that the field space trajectory is primarily along the $\varphi$ direction. In contrast, in between the two turns in field space (second phase), $\varepsilon_{\varphi} \gg \varepsilon_{\chi}$ and the evolution is predominantly along the $\chi$ direction. We call this intermediate evolution the ``$\varepsilon$-floor'' phase, and highlight it in yellow in Fig.~\ref{fig:ILL-bckgr} and in subsequent figures.

\section{Perturbations}
\label{sec:pert}

We now study the evolution of cosmological perturbations in the Fibre Inflation PBH scenario. We first establish the formalism, following Refs.~\cite{Sasaki:1995aw,Gordon:2000hv,Wands:2002bn,Langlois:2008mn,Peterson:2010np,Gong:2011uw,Kaiser:2012ak,Gong:2016qmq}, and then apply it to the model of interest.

\subsection{Formalism}
\label{sec:pert:form}

We decompose each of the fields into a background and a fluctuation,
\begin{equation}
    \phi^I(x,t) = \phi^I(t) + \delta \phi(x,t) \,,
\end{equation}
and from this construct the gauge-invariant Mukhanov-Sasaki variables,
\begin{equation}\label{eq:MS-var-def}
    Q^I \equiv \delta\phi^I + \frac{\dot{\phi}^I}{H}\psi\, ,
\end{equation}
where $\psi$ is the scalar metric perturbation on equal-time hypersurfaces and $I=1,\dots,\calN$ labels the fields. 

We decompose perturbations into {\it adiabatic} perturbations, which are parallel to the field-space trajectory and {\it isocurvature} perturbations that are perpendicular to it. The former are defined by the unit vector $\hat{\sigma}^I$, given by Eq.~\eqref{eq:sigma_hat}, while the latter correspond to perturbations along the $\calN-1$ directions spanned by
\begin{equation}
    \hat{s}^{IJ} \equiv {\cal G}^{IJ} - \hat{\sigma}^I \hat{\sigma}^J \,.
\end{equation}
The Mukhanov-Sasaki field variables can be projected along the adiabatic and isocurvature fluctuations as
\begin{equation}\label{eq:MS-var-decomp}
    Q^I = \hat{\sigma}^I Q_\sigma + \hat{s}^I_{\,J}\, Q_s ^J\,,
\end{equation}
or equivalently,
\begin{equation}\label{eq:QsigmasI}
    Q_\sigma \equiv \hat{\sigma}_I Q^I \,, \qquad
    Q_s ^I \equiv \hat{s}^I_{\>\> J} Q^J \,.
\end{equation}
From this we construct the canonically normalized comoving adiabatic and isocurvature perturbations as
\begin{equation}\label{eq:R-and-S-def}
    {\cal R} = \frac{H}{\dot{\sigma}}Q_\sigma \,, \qquad 
    {\cal S}^I = \frac{H}{\dot{\sigma}}Q_s  ^I \,.
\end{equation}
Note that while $I$ takes on values from $1$ to $\calN$, only $\calN-1$ components of ${\cal S}^I$ are linearly independent \cite{Kaiser:2012ak}.

For simplicity, in what follows we focus on the impact of each axion field separately. In this two-field limit, there is only a single independent isocurvature perturbation, which we denote by ${\cal S}$.
The evolution of the Fourier modes of the curvature and isocurvature perturbations is governed by \cite{Kaiser:2012ak,Gong:2016qmq,Wands:2007bd,Achucarro:2010da,Achucarro:2016fby,McDonough:2020gmn,Lorenzoni:2024krn}
\begin{gather}
    \frac{d}{dt}  \left(\dot {\cal R}_k - 2 \omega {\cal S}_k \right)
    + (3 + \delta) H \left( \dot  {\cal R}_k - 2 \omega {\cal S}_k \right)  + \frac{k^2}{a^2}{\cal R}_k = 0\,,
    \label{eq:EoM-R}
    \\
    \ddot{\cal S}_k+ (3+ \delta) H \dot{\cal S}_k + \left(\frac{k^2}{a^2} + \mu^2_s\right){\cal S}_k = - 2 \omega \dot{\cal R}_k\,,
    \label{eq:EoM-S}
\end{gather}
where $\delta= 4\varepsilon-2\eta$ and $\omega$ is the pseudo-scalar turn rate, defined in the two-field case as \cite{McDonough:2020gmn}
\begin{equation}
    \omega\equiv \sqrt{\det \calG_{IJ}}\,\epsilon_{IJ}\,\hat{\sigma}^I \omega^J\,,
\end{equation}
with the usual Levi-Civita symbol $\epsilon_{IJ}$; in this limit, $\omega=\pm|\omega^I|$.
In Eq.~\eqref{eq:EoM-S}, $\mu_s^2$ acts as an effective isocurvature mass squared,
\begin{equation}\label{eq:muS}
    \mu_s ^2 = {\cal M}_{ss} - {\cal M}_{\sigma\sigma} + 2 H^2 \varepsilon\left(3+\delta-\varepsilon\right) \,,
\end{equation}
with ${\cal M}$ being the perturbations' mass matrix with components in the field basis \cite{Kaiser:2012ak},
\begin{equation}\label{eq:MIJ}
    {\cal M}^I_{\>\> J} \equiv {\cal G}^{IK} \left( {\cal D}_J {\cal D}_K V \right) - {\cal R}^I_{\>\> LMJ} \dot{\phi}^L \dot{\phi}^M \,.
\end{equation}
Here ${\cal D}_I$ denotes a covariant derivative with respect to the field space metric and ${\cal R}^I_{\>\> LMJ}$ is the Riemann tensor for the field-space manifold. The relevant components in the adiabatic and isocurvature basis are given by
\begin{eqnarray}
    {\cal M}_{\sigma\sigma} &&\equiv \hat{\sigma}_I \hat{\sigma}^J {\cal M}^I_{\>\> J} = \hat{\sigma}^K \hat{\sigma}^J \left( {\cal D}_K {\cal D}_J V \right)\,,\\
    {\cal M}_{ss} &&\equiv \hat{s}_I\, \hat{s}^J\, {\cal M}^I_{\>\> J}= \hat{s}^K\, \hat{s}^J\, \left( {\cal D}_K {\cal D}_J V \right) - \varepsilon H^2 \mathbb{R}\,,
\end{eqnarray}
where $\hat{s}^I \equiv \omega^I / \omega $, and we note that the Riemann tensor drops out of ${\cal M}_{\sigma\sigma}$ but not from ${\cal M}_{ss}$: in fact, $\mathbb{R}$ is the Ricci scalar of the field-space manifold.

The equations of motion simplify on long wavelengths, to
\begin{gather}
    \dot {\cal R}_k \simeq
    2\omega \calS_k\,,
    \label{eq:EoM-R-superHubble}
    \\
    \ddot{\cal S}_k+ (3+ \delta) H \dot{\cal S}_k + \tilde{\mu}^2_s{\cal S}_k \simeq 0\,,
    \label{eq:EoM-S-superHubble}
\end{gather}
where the $\tilde{\mu}_s^2$ is the effective super-horizon isocurvature mass squared, defined by
\begin{equation}
  \tilde{\mu}_s^2 \equiv  \mu_s^2 + 4\omega^2 \,\, \,\, (\text{super-horizon})\,.
  \label{eq:tildemus}
\end{equation}
This system of equations allows for the dynamical evolution of modes far outside the horizon, including a sourcing of ${\cal R}$ from $\calS$ whenever the turn rate $\omega$ is non-vanishing.

\subsection{Evolution of Perturbations}
\label{sec:pert:evol}

We now apply this formalism to the Fibre Inflation PBH scenario. The three-phase structure of the background evolution (see Fig.~\ref{fig:ILL-bckgr}) implies that the correspondence between $\{ {\cal R}, {\cal S}\}$ and the field fluctuations $\{ \delta\varphi,\delta\chi\}$ changes over time. At early and late times, when $\varepsilon_{\varphi} \gg \varepsilon_\chi$, inflation proceeds along the $\varphi$ direction, and ${\cal R}$ and ${\cal S}$ correspond to fluctuations along the $\delta \varphi$ and $\delta \chi$ directions, respectively. These roles are reversed in the intermediate ``$\varepsilon$-floor'' phase, when $\varepsilon_{\chi} \gg \varepsilon_\varphi$: inflation proceeds along the $\chi$ direction, and ${\cal R}$ and ${\cal S}$ now correspond to fluctuations along the $\delta \chi$ and $\delta \varphi$ directions, respectively.

\begin{figure}[htb!]
    \centering
    \includegraphics[width=\linewidth]{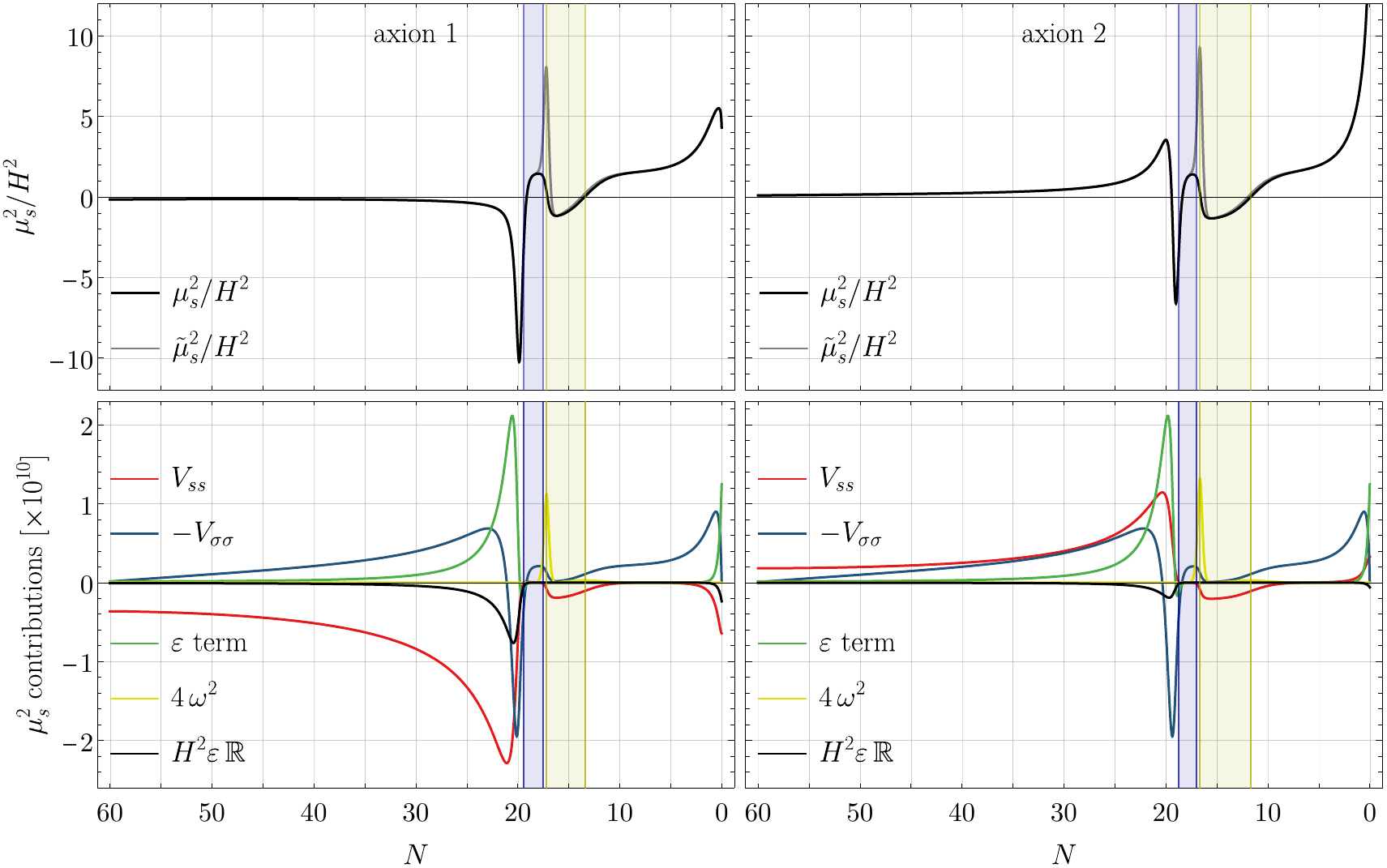}
    \caption{Evolution of the effective isocurvature mass squared and its components for the axion-1 ($\chi_1$) (\textit{left panels}) and axion-2 ($\chi_2$) (\textit{right panels}) systems, with parameters given in Tab.~\ref{tab:ILL-params}. 
    \textit{Top panels:} effective isocurvature mass squared for sub-Hubble (black) and super-Hubble (gray) perturbations.
    \textit{Bottom panels:} components of $\tilde\mu_s^2$, from Eq.~\eqref{eq:muS-components}.}
    \label{fig:muS}
\end{figure}

We first examine the evolution of the isocurvature mass squared, $\mu_s^2$, for which the phase structure of the background has important consequences. This is shown in Fig.~\ref{fig:muS}, which shows the isocurvature mass squared, Eq.~\eqref{eq:muS}, in black and the effective super-horizon isocurvature mass squared, Eq.~\eqref{eq:tildemus}, in gray. From Fig.~\ref{fig:muS}, one may appreciate that during the ``$\varepsilon$-floor'' phase (yellow shading) both axion models have $\mu_s^2<0$, indicating that they exhibit a {\it tachyonic} isocurvature.
In this phase, the mass-matrix projections become $\calM_{\sigma\sigma}\approx\pd_\chi^2 V_{\rm ax}$ and $\calM_{ss}\approx\pd_\varphi^2 V_{\rm inf}$, and the effective isocurvature mass squared simplifies to
\begin{equation}
\label{eq:musapprox}
    \mu_s^2\approx \pd_\varphi^2 V_{\rm inf}(\varphi) - \pd_\chi^2 V_{\rm ax}(\chi)\, ,
\end{equation}
since the last term in Eq. \eqref{eq:muS} is suppressed by $\varepsilon$.  The first term in Eq.~\eqref{eq:musapprox} must become negative in order for $\varphi$ to emerge from the plateau feature in $V_{\rm inf}$, while the second term in Eq.~\eqref{eq:musapprox} is negative definite if $\chi/f < \pi$, which is the case for these parameter sets. This leads to a tachyonic instability of the isocurvature perturbations, $\mu_s^2<0$, during most of the ``$\varepsilon$-floor'' phase, for both the axion-1 ($\chi_1$) and axion-2 ($\chi_2$) models. 

The axions differ in the evolution of the isocurvature mass squared at early times. The axion $\chi_1$ already exhibits a tachyonic isocurvature during the first phase of inflation. To see this, we may express the effective super-horizon isocurvature mass squared, Eq.~\eqref{eq:tildemus}, as
\begin{equation}\label{eq:muS-components}
    \tilde\mu_s^2= V_{ss}- V_{\sigma\sigma}+2\,\varepsilon \,H^2(3+\delta-\varepsilon)+4\omega^2+\varepsilon\, H^2 \mathbb{R}\,,
\end{equation}
where we defined $V_{ss}\equiv \hat{s}^I \hat{s}^J \calD_{I} \calD_J V$ and $V_{\sigma\sigma}\equiv \hat{\sigma}^I \hat{\sigma}^J \calD_I \calD_J V$. Note that the field-space Ricci scalar $\mathbb{R}$ is negative in both cases of $\chi_1$ and $\chi_2$. As illustrated in the bottom panel of Fig.~\ref{fig:muS}, the tachyonic evolution of $\tilde\mu_s^2$ for $\chi_1$ is driven by the term $V_{ss}<0$. Using Eq.~\eqref{eq:covariant-derivative}, this quantity is itself given by
\begin{equation}\label{eq:VNN}
    V_{ss}=\hat{s}^I \hat{s}^J(\pd_I\pd_J V-{\Gamma}^K_{\,IJ}\pd_K V)\,.
\end{equation}
At early times, inflation proceeds along the $\hat{\sigma}\approx \hat{\varphi}$ direction and hence $\hat{s}\approx \hat{\chi}$ and $V_{ss}\approx \partial_\chi^2\,V - \Gamma^K _{\chi \chi} \partial_K V$.
The only non-vanishing Christoffel symbol is
\begin{equation}
    {\Gamma}^\varphi_{\,\chi\chi}= 2\,b\,\e^{-2b\varphi}\,,
\end{equation}
where $b=+\frac{2}{\sqrt{3}}$ for $\chi_1$ and $b=-\frac{1}{\sqrt{3}}$ for $\chi_2$. We note also that $\pd_\chi^2\,V\approx 0$ (see bottom panels), and hence conclude that $V_{ss}<0$ due to the sign of the kinetic coupling: positive for $\chi_1$ and negative for $\chi_2$, leading to a tachyonic isocurvature mass for the former.

\begin{figure}[ht!]
    \centering
    {PBH-scale perturbations}
    \includegraphics[width=\linewidth]{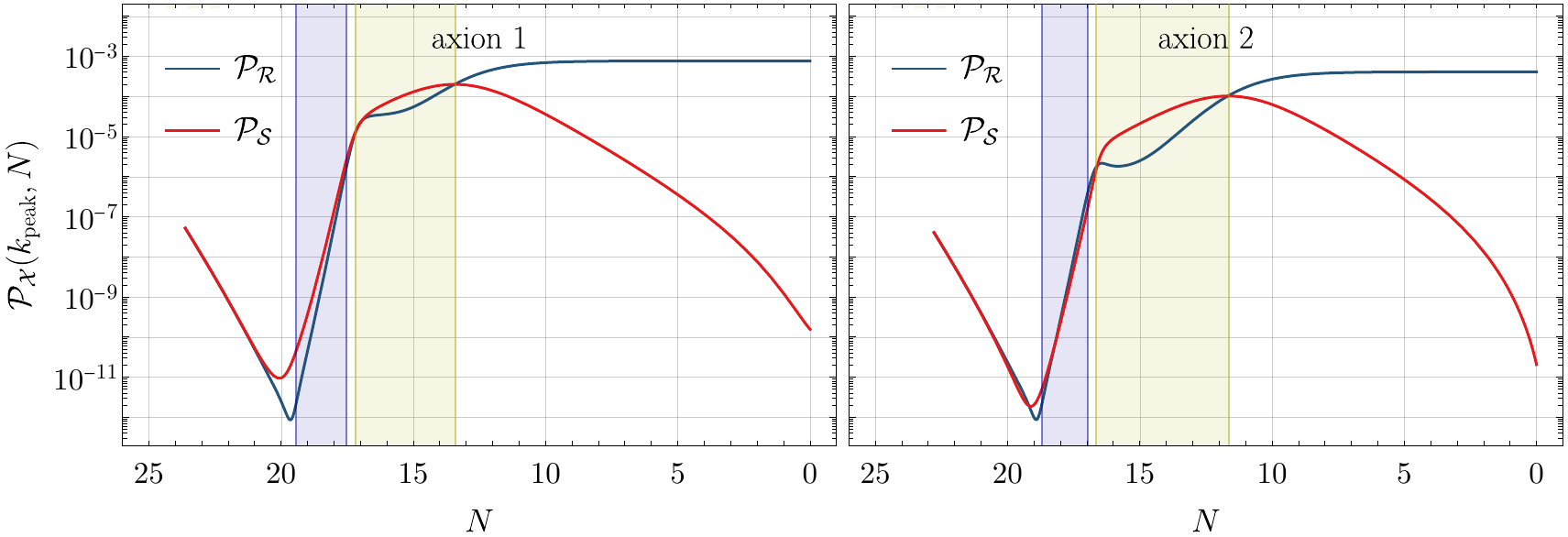}
    \caption{
    Evolution of PBH-scale curvature and isocurvature perturbations $\calR_{k_\text{peak}}$ and $\calS_{k_\text{peak}}$ in the Fibre Inflation PBH model, with a dynamical axion $\chi_1$ ({\em left panel}) and axion $\chi_2$ ({\em right panel}). We consider the ``Axion-1'' and ``Axion-2'' parameters in Tab.~\ref{tab:ILL-params}.}
    \label{fig:PBH-perturbations}
\end{figure}

\begin{figure}[ht!]
    \centering
    {CMB-scale perturbations}
    \includegraphics[width=\linewidth]{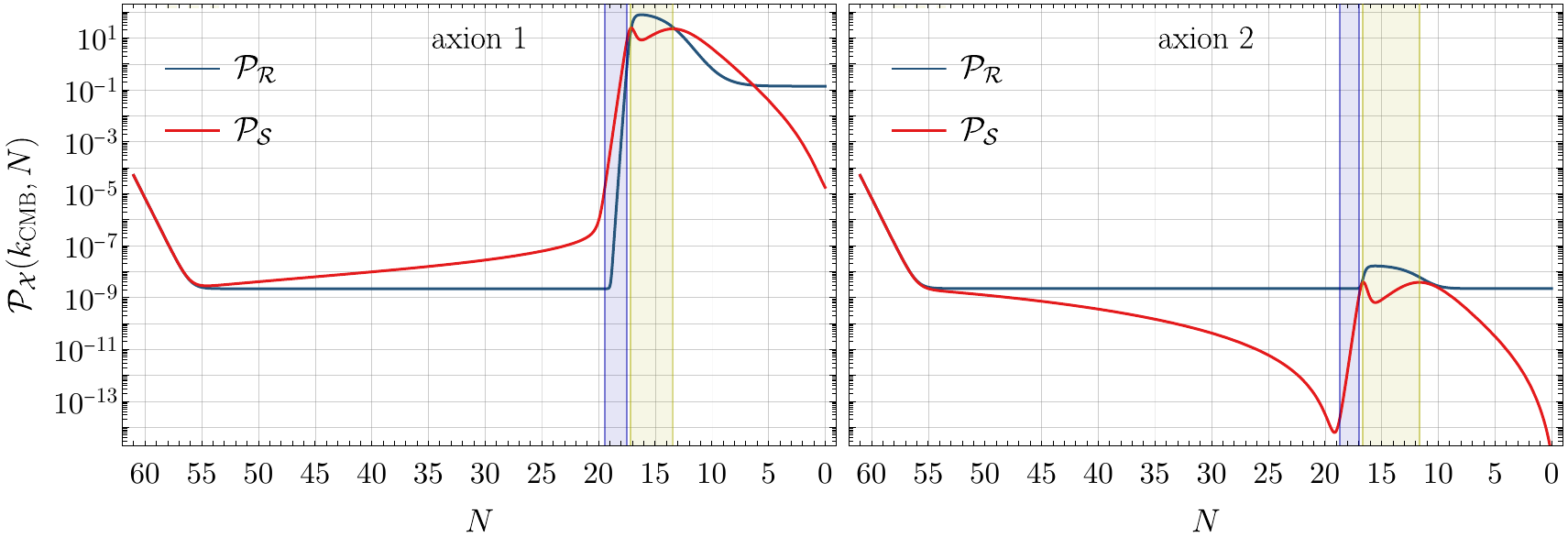}
    \caption{
    Evolution of CMB-scale curvature and isocurvature perturbations $\calR_{k_\text{CMB}}$ and $\calS_{k_\text{CMB}}$ in the Fibre Inflation PBH model, with a dynamical axion $\chi_1$ ({\em left panel}) and axion $\chi_2$ ({\em right panel}). We consider the ``Axion-1'' and ``Axion-2'' parameters in Tab.~\ref{tab:ILL-params}.}
    \label{fig:CMB-perturbations}
\end{figure}

We now numerically solve the perturbation equations, Eq.~\eqref{eq:EoM-R} and Eq.~\eqref{eq:EoM-S}, for the ``Axion-1'' and ``Axion-2'' parameter sets of Tab.~\ref{tab:ILL-params}. The results are shown in Fig.~\ref{fig:PBH-perturbations} and Fig.~\ref{fig:CMB-perturbations}, which show the evolution of modes $k$ on scales relevant to PBHs and scales that are probed by CMB data, respectively. We consider the axions separately and provide illustrative examples. Results for the power spectrum of curvature perturbations will be discussed in Sec.~\ref{sec:power-spectrum}. For reference, the background evolution for these parameter sets is shown in Appendix~\ref{app:params}, Figs.\ref{fig:ax1-bckgr} and~\ref{fig:ax2-bckgr}. 

Fig.~\ref{fig:PBH-perturbations} shows the evolution of PBH-scale modes of ${\cal R}$ and ${\cal S}$ in the models with $\chi_1$ (left) and $\chi_2$ (right). In both cases, $\chi_1$ and $\chi_2$, the evolution of perturbations inherits the three-phase structure of the background evolution. Both ${\cal R}$ and ${\cal S}$ exit the horizon before the first field-space turn, and grow during the phase of ultra-slow roll inflation indicated by the blue-shaded band. Subsequently, during the ``$\varepsilon$-floor'' phase indicated by the yellow-shaded band, the isocurvature becomes tachyonic, which enhances $\calS$ beyond the level already produced by USR. As the ``$\varepsilon$-floor'' phase ends and the trajectory turns back towards the inflaton direction (i.e. $\omega\neq0$), the amplified ${\cal S}$ sources a curvature perturbation ${\cal R}$. At late times, the isocurvature becomes heavy and decays, while the curvature freezes-out to its amplified value.

This can be compared with the evolution of CMB-scale modes, which is shown in Fig.~\ref{fig:CMB-perturbations}. The two axions now stand in stark contrast: $\chi_1$ amplifies the curvature perturbation on CMB scales, while $\chi_2$ leaves $\calR_{k_{\rm CMB}}$ at late times unchanged from its value at horizon exit. 
The difference between the axions derives from the isocurvature mass squared, which is negative even at early times for $\chi_1$, driving an early growth of isocurvature perturbations; this persists to late times, with $\chi_1$ producing a significant enhancement of $\calR_{k_{\rm CMB}}$. In contrast, $\chi_2$ shows a transient sourcing of ${\cal R}$ when the trajectory turns towards the axion direction, which is excised when the trajectory reverts back to the inflaton direction (meaning $\omega$ will have an opposite sign with respect to the first turn), restoring ${\cal R}$ to its value at horizon exit.

\section{Imprint on the Power Spectrum}
\label{sec:power-spectrum}

We now present results for the power spectrum of the comoving curvature perturbation, in the case of $\chi_1$, $\chi_2$, and jointly. We first emphasize the central role played by the decay constant: it determines not only the dynamical evolution of the axion mass and kinetic terms during inflation, as shown in Fig.~\ref{fig:axion-phi-dependence}, but also the overall normalization of the mass, Eq.~\eqref{eq:axion-mass-0}, which is exponentially sensitive to the decay constant, $m_{\rm ax}^2\propto e^{-\mpl/f}$. Axions which are very light relative to the Hubble parameter during the whole inflationary evolution, $m_{\rm ax}\ll H$ at all times, have no impact on the power spectrum of curvature perturbations, and hence do not impact PBH production. Consistent with this, we find that an axion with a decay constant much less than the Planck scale, $f\ll \mpl$, has no impact on the power spectrum. 

\begin{figure}[ht!]
    \centering
    \includegraphics[width=\linewidth]{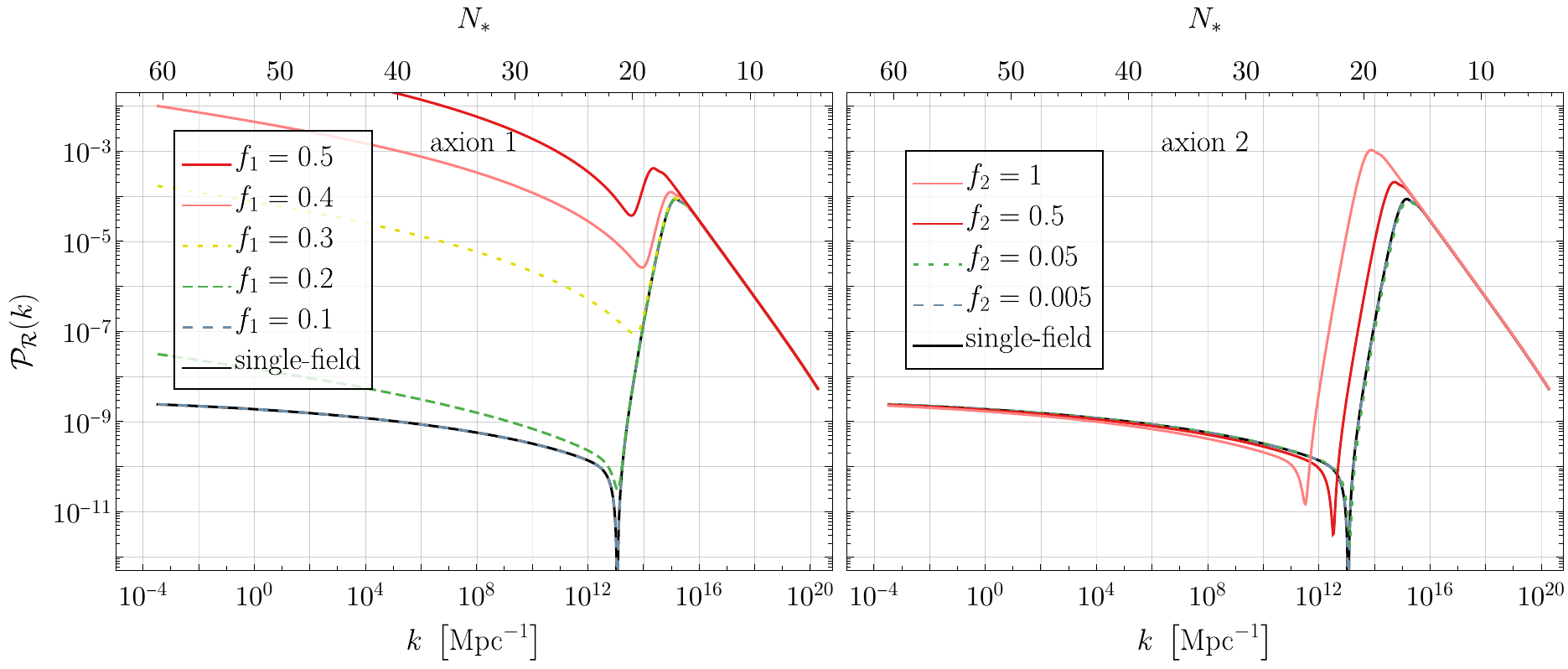}
    \caption{
    Dependence of the curvature power spectrum on the value of the axion decay constant, for $\chi_1$ (\textit{left panel}) and $\chi_2$ (\textit{right panel}).
    }
    \label{fig:axion-fa-scan}
\end{figure}

This is illustrated in Fig.~\ref{fig:axion-fa-scan}, which shows the power spectrum of curvature perturbations for the case of a dynamical axion $\chi_1$ (left panel) and axion $\chi_2$ (right panel), for varying values of the axion decay constant. We fix the axion mass using Eq.~\eqref{eq:axion-mass-0} with $|W_0|=1$, ${\cal V}=10^{3}$, $A_1=A_2=10^{-6}$, and varying $f_1$ and $f_2$. Such a small value of the prefactor of the non-perturbative effects is needed, in particular, for $A_2$ to ensure that the axion potential remains always subdominant with respect to the inflaton one. In fact, as shown in Sec. \ref{sec:axions}, the mass of $\chi_2$ becomes an order of magnitude smaller than the Hubble scale during the USR phase only if $f_2 \gtrsim 0.1\,\mpl$. However, as can be seen from (\ref{eq:axion-mass-0}), for such a large value of $f_2$, the exponential suppression $e^{-1/f_2}$ in $M_2^2$ is not very effective, requiring $A_2\ll 1$ to keep the axion $\chi_2$ light enough. Smaller values of $f_2$ might seem to be compatible with $\mathcal{O}(1)$ values of $A_2$ but, in that case, the inflaton-dependent axion mass in (\ref{eq:axion-mass}) would always remain much smaller than $H$. This tuning in $A_2$ is a crucial condition to realize axion-assisted PBH formation. Note that in the case of $\chi_1$, one could instead reproduce a similar dynamics for smaller values of $f_1$ and $\mathcal{O}(1)$ values of $A_1$.

One may appreciate in Fig.~\ref{fig:axion-fa-scan} that for $f_{1,2} \lesssim 0.1 \,\mpl$, the axions have no impact on the power spectrum. Given the many orders of magnitude of uncertainty on the decay constants, from this we conclude that the Fibre Inflation PBH model is robust to the inclusion of the axions for generic axion parameters.

On the other hand, axion fields with mass comparable to Hubble ($m_{\rm ax}\simeq {\cal O}(0.1) H$) at some point during inflation, and decay constant near but sub-Planckian ($f\gtrsim 0.1 \,\mpl$) can have a significant impact on the power spectrum. In the case of $\chi_1$, the relative enhancement of CMB modes is completely degenerate with an overall renormalization of the potential, which obviates the modification to CMB scale but also removes any relative enhancement on PBH scales. Meanwhile, $\chi_2$ serves to modify the power spectrum on PBH scales while leaving CMB scales unchanged. For $f_2\gtrsim 0.3 \,\mpl$, the peak of the power spectrum can be enhanced by up to one order of magnitude. We point out that, for a small region of parameter space around $f_2\sim0.2\,\mpl$, $\chi_2$ can instead slightly suppress the peak of $\prk(k)$; this is due to a decreasing mass coupling at those times, see Fig.~\ref{fig:axion-phi-dependence}.

Let us point out that values of $f_2\gtrsim 0.3 \,\mpl$ require a very large rank of the condensing gauge group. In fact, matching the observed amplitude of density perturbations in Fibre Inflation typically sets $\langle\mathcal{V}\rangle\simeq \sqrt{\langle\tau_1\rangle}\langle\tau_2\rangle\sim \mathcal{O}(10^3)$. Requiring an inflaton field range inside the K\"ahler cone large enough to yield at least $N_e\gtrsim 52$ e-foldings of inflation then generically gives $\langle\tau_1\rangle\sim \mathcal{O}(5)$ and $\langle\tau_2\rangle\sim\mathcal{O}(500)$. Plugging this value inside the definition of the decay constant $f_2$ in Eq. (\ref{eq:fa-def}), one then finds that $f_2\gtrsim 0.3 \,\mpl$ correlates with $N_2\gtrsim\mathcal{O}(10^3)$, which demands a substantial effort in model building.

Let us stress, however, that $f_2\gtrsim 0.3\,\mpl$ is perfectly compatible with a controlled effective field theory. In fact, one might worry that, for such a large $f_2$, the condition $a_2\tau_2\gg 1$ cannot be satisfied, since $a_2 \tau_2=\frac{\tau_2}{f_2\langle\tau_2\rangle}\lesssim 3.3\,\frac{\tau_2}{\langle\tau_2\rangle}$ becomes smaller than unity at the onset of inflation where $\tau_2\ll\langle\tau_2\rangle$. However, the requirement $a_2\tau_2\gg 1$ applies only when the superpotential is generated by a series of string instanton corrections and higher-order contributions are neglected. In our case, we instead consider gaugino condensation, which is an exact IR non-perturbative effect and does not require the condition $a_2\tau_2\gg 1$ to hold. The only requirement for trusting gaugino condensation is $\tau_2\gg 1$, ensuring a weakly coupled gauge theory. This condition is always satisfied in our case even when $a_2\tau_2< 1$ due to $a_2=2\pi/N_2\ll 1$ for $N_2\sim\mathcal{O}(10^3)$.

\begin{figure}
    \centering
    \includegraphics[width=0.6\linewidth]{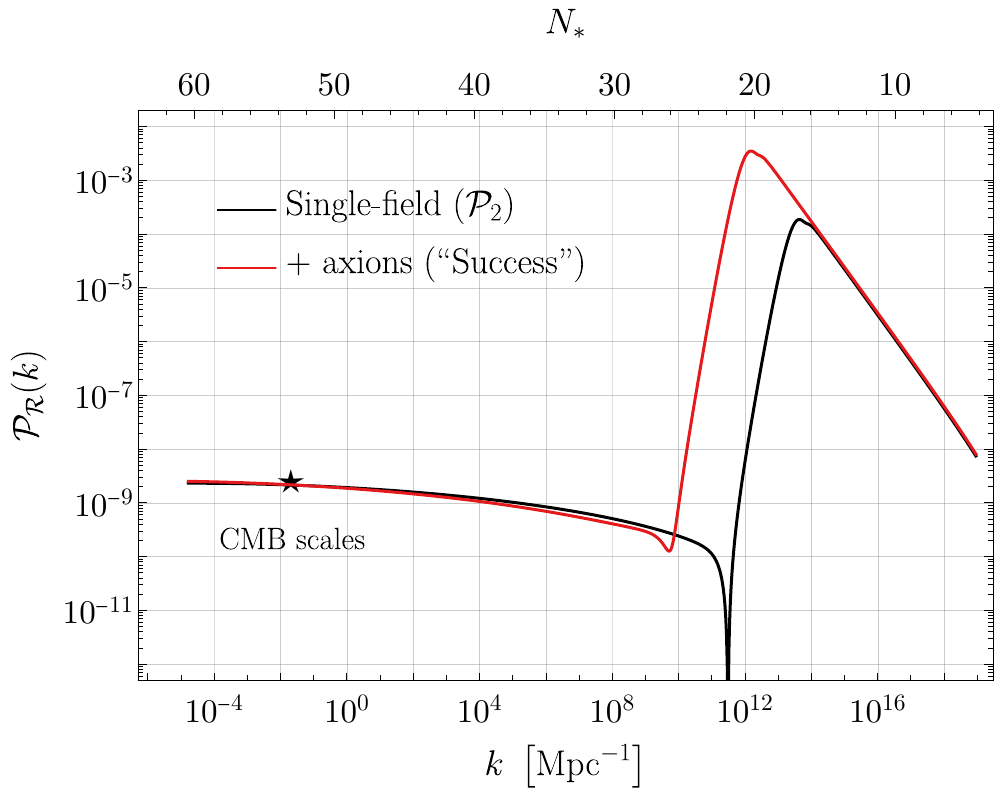}
    \caption{
    Curvature power spectrum for the ``Success'' model with two dynamical axions, compared to its single-field limit. The axions contribute to enhancing the peak of $\prk(k)$, enabling the production of primordial black holes.}
    \label{fig:PRk-success}
\end{figure}

Fig.~\ref{fig:PRk-success} provides an example where both axions are dynamical during inflation (see Tab. \ref{tab:SUC-params} for the corresponding parameter values). This example is compatible with CMB observations (given in Tab.~\ref{tab:SUC-obs}) and features a significant enhancement of the peak of the power spectrum. Indeed, the underlying single-field model (black) falls well below the threshold for PBH production (${\cal P}_{\cal R}(k) \gtrsim 10^{-3}$), while the model with axions (red) is well above the threshold. This indicates that dynamical axions can assist the production of primordial black holes from inflation in string theory.

\begin{table}[ht!]
    \centering
    \begin{tabular}{@{\extracolsep{5pt}}c|*{6}c}
       \textbf{Model} & $M_1$ & $f_1$ & $\chi_{1,i}$ & $M_2$ & $f_2$ & $\chi_{2,i}$ \\ \hline
       \textbf{Success} & $8\times10^{-7}$  & $0.1$ & $0.1\pi f_1$ & $8\times10^{-7}$ & $0.5$ & $0.1 \pi f_2$ \\
    \end{tabular}
    \caption{Parameter values for the axionic potential~\eqref{eq:V-ax}, for the ``Success'' axion model. The parameters $M_{1,2}$ represent the axion masses around the minimum of $V_{\rm inf}$, as defined in Eq.~\eqref{eq:axion-mass-0}.
    This model uses the $\calP_2$ parameters of Tab.~\ref{tab:SF-params} for the base inflationary potential~\eqref{eq:V-inf-new}, with the rescaling of $V_0=2.07\times 10^{-9}$ to match CMB observables. 
    }
    \label{tab:SUC-params}
\end{table}

\begin{table*}[htb!]
    \centering
    \makebox[\textwidth][c]{
    \begin{tabular}{@{\extracolsep{-1pt}}c|*{7}c|*{1}c}
        \textbf{Model}
        & $A_s$ & $n_s$ & $\alpha_s$ & $r$ & $\beta_{\rm iso}$ & $f_{\rm NL}^{\rm equil}$ & $M_{\rm PBH}~
        [{\rm g}]$
        & $\prk^{\rm peak}$ 
        \\ 
        \hline 
        \textbf{Success}
        & $2.10\times 10^{-9}$ & $0.9653$ & $-0.003$ & $0.002$ & $0.0001$ & $0.91$ & $2.5\times 10^{21}$
        & $3.5\times 10^{-3}$ 
    \end{tabular}
    }
    \caption{Observables for ``Success'' model as defined in Tab.~\ref{tab:SUC-params}. CMB scales exit at $N_{\rm CMB} = 53$; observables are computed using \texttt{PyTransport}~\cite{Mulryne:2016mzv,Ronayne:2017qzn}.
    This model successfully enhances the peak of the curvature power spectrum with respect to the single-field limit $\calP_2$, thus enabling production of PBHs.
    }
    \label{tab:SUC-obs}
\end{table*}

\section{Discussion and Conclusions}
\label{sec:disc}

Primordial black holes provide a window into the microphysics of cosmic inflation, and may reveal clues towards a string theory origin of the Universe. In this work, we have studied a concrete model of inflation in string theory capable of satisfying cosmic microwave background constraints on large scales while providing the large enhancement of small-scale perturbations needed to seed primordial black holes. Our analysis has focused on the role that string theory axions might play in this scenario. 

Our results indicate that Fibre Inflation provides a robust mechanism for the production of PBHs. The model can accommodate axions with generic model parameters without significant alteration to the observational predictions. We also find a tuned region of parameter space wherein axions provide a helping hand, by boosting the amplification of perturbations already present in the single field model. In particular, Fig.~\ref{fig:PRk-success}, red curve, shows a fiducial example of Fibre Inflation with two dynamical axions satisfying CMB constraints and producing PBHs, while the underlying single field model (black), shown in black, falls well below the threshold for PBH formation.

There are many directions for future work. On the observational side, the model is testable through gravitational waves. Not only through the tensor-to-scalar ratio, which is within reach of next generation CMB experiments, or the scalar-induced gravitational waves (studied in Fibre Inflation in Ref.~\cite{Cicoli:2022sih}), but also by the gravitational waves sourced by the axions, e.g.~, indirectly by the axion coupling to gauge fields \cite{Barnaby:2011qe}, or directly by the axion coupling to gravity as in Ref.~\cite{Alexander:2004us}. The model therefore predicts a {\it triplet} of gravitational wave signatures: the vacuum fluctuations from inflation, the scalar-induced gravitational waves from the peak in the scalar power spectrum, and those sourced by the axions.

On the model building side, further work is required to fully understand whether globally consistent K3-fibred Calabi-Yau compactifications exist that allow for gaugino condensation on a stack of $N_2\sim\mathcal{O}(10^3)$ branes, leading to successful Fibre Inflation with an axion decay constant $f_2\gtrsim 0.3\,\mpl$ which enables axion-assisted PBH formation. The major challenge is definitely expected to be tadpole cancellation which sets an upper bound on the number of D7-branes. Finding type IIB examples with $N_2\sim\mathcal{O}(10^3)$ is therefore challenging but not impossible since F-theory compactifications are known to be able to lead to very large ranks. For example, Ref. \cite{Candelas:1997pq} presented an F-theory compactification on a 4D compact fourfold where the largest gauge group is $SO(7232)$ that would indeed lead, if condensing, to the required value of the axion decay constant.

Moreover, we have seen that axion-assisted PBH formation requires an exponentially small prefactor of the non-perturbative effects generating the axion potential, $A_2\simeq 10^{-6}$, while these coefficients are expected to be $\mathcal{O}(1)$. Two possible ways to achieve such a small $A_2$ could be large threshold corrections to the gauge kinetic function, or a field-dependent prefactor with a small field VEV. In the first case the gauge kinetic function $\mathfrak{f}_2=T_2$ would receive corrections of the form $\mathfrak{f}_2=T_2+\Delta$ which, when inserted into $W\supset A_2\,e^{-2\pi\mathfrak{f}_2/N_2}$, would generate an effective prefactor, $A_{{\rm eff},2} = e^{-2\pi{\rm Re}(\Delta)/N_2}$, that could be exponentially small if $2\pi{\rm Re}(\Delta)>N_2$. This condition is, however, challenging to realize while keeping control over the perturbative expansion which requires ${\rm Re}(\Delta)<\tau_2$. The second case seems instead more promising since it exploits the fact that $A_2$ can depend on complex structure moduli or matter fields. Indeed, $A_2$ could be close to zero near special loci in complex structure moduli space. Alternatively, symmetries might force $A_2$ to depend on a charged matter field $\phi$ as $A_2\sim \phi^n$, which could lead to a tiny prefactor for $\phi\ll 1$ in Planck units.

Moreover, another feature of this model that deserves closer examination is the transition from inflation to radiation domination, namely the epoch of reheating. Reheating probes the global structure of the compactification, and can impose non-trivial constraints on the allowed phenomenology, as studied in Ref. \cite{Cicoli:2018cgu,Cicoli:2022uqa} which studied reheating in Fibre Inflation from the decay of the inflaton with Standard Model degrees of freedom on either D7- or D3-branes. In models with axion-assisted PBH formation, as explained above, the axion masses are only about one order of magnitude below the Hubble scale during inflation. As a result, the axions are sufficiently heavy to decay during the post-inflationary evolution. Given that reheating is driven by the decay of the longest-lived fields, we expect reheating in these models to be governed, not by the decay of the inflaton, but by the decay of the light axions into gauge fields and fermions. 

These considerations depend crucially on the precise values of the axion masses, which, in turn, depend on the order in the non-perturbative expansion at which the axionic shift symmetry is broken. In this work, in order not to make the axions too light, we assumed this breaking to occur at leading non-perturbative order via a contribution to the superpotential from gaugino condensation. However, K3-fibred Calabi-Yau threefolds are characterized by the presence of non-rigid cycles, which in general lead to matter fields in the adjoint representation which can forbid gaugino condensation, contrary to the typical case of a  pure SYM theory. Non-perturbative effects could then lead to a highly suppressed axion potential at multi- or poly-instanton order \cite{Blumenhagen:2012kz,Caraffi:2026wzk}. It would therefore be interesting to investigate if these additional adjoint zero modes can be lifted through a globally consistent choice of fluxes, thereby realizing explicitly the non-perturbative effects required for successful axion-assisted PBH formation.

Finally, it would be interesting to connect this analysis with the larger body of work on phenomenology of fibred Calabi-Yau threefolds, such as dark energy model building \cite{Cicoli:2012tz,Cicoli:2021skd,Cicoli:2024yqh,Cicoli:2026bqo}. It may be that one of the axions studied here can assist PBH dark matter while the other can provide a realization of quintessence. We leave this, and other interesting directions, to future work.

\section*{Acknowledgments}

The authors thank Naman Agarwal, Andrew Frey, Sarah Geller, and David Kaiser for helpful discussions. 
E.M. is supported in part by a Discovery Grant from the Natural Sciences and Engineering Research Council of Canada, and by a New Investigator Operating Grant from Research Manitoba. 
D.L. is supported in part by a PhD Research Studentship from Research Manitoba, and by a Canada Graduate Research Scholarship -- Doctoral from the Natural Sciences and Engineering Research Council of Canada (NSERC) [funding reference number CGRS D - 611992].

\appendix
\section{Models considered}
\label{app:params}

In what follows, we summarize all the models considered in the previous sections, present their parameters and their observables, and show the background evolution when not present in the main body. Six models were presented throughout this work: 
\begin{itemize}
    \item The ``Reference'' single-field model $\calP_1$ reproduces the results of the homonymous model in Ref.~\cite{Cicoli:2022sih}; it is the subject of Sec.~\ref{sec:pbh}.
    
    \item The ``No-PBH'' single-field model $\calP_2$ modifies the $C_i$ parameters such as to produce a less prominent enhancement of the curvature power spectrum on small scales, insufficient for producing primordial black holes, while maintaining consistency with CMB observables; it is introduced in Sec.~\ref{sec:pbh}, and provides the basis for all of the following axion models, which all adopt the $\calP_2$ parameters for $V_{\rm inf}$.
    
    \item The ``Illustrative'' three-field model highlights the impact that the two axions of Eq.~\eqref{eq:action-full} have on the inflationary dynamics; it is the subject of Sec.~\ref{sec:axions}.
    
    \item The ``Axion-1'' two-field model highlights the impact of the axion $\chi_1$ on the inflationary dynamics; it is considered in Sec.~\ref{sec:pert}.
    
    \item The ``Axion-2'' two-field model highlights the impact of the axion $\chi_2$ on the inflationary dynamics; it is considered in Sec.~\ref{sec:pert}.
    
    \item The ``Success'' three-field model provides a physical example wherein the two dynamical axions enhance the peak of the curvature power spectrum, making the production of PBHs possible while matching CMB observables; it is considered in Sec.~\ref{sec:power-spectrum}.
\end{itemize}
We explicitly report all the parameter values considered for each of these models in Tab.~\ref{tab:all-params}. The corresponding observables are summarized in Tab.~\ref{tab:all-obs} for the single-field and ``Success'' models (the illustrative models are not meant to be physical).

\begin{table}[ht!]
    \centering
    \begin{tabular}{@{\extracolsep{0pt}}c|*{6}c}
        \shortstack{\textbf{Model}\\~} & 
        \shortstack{\textbf{Reference}\\$\boldsymbol{({\cal P}_1)}$} & 
        \shortstack{\textbf{No-PBH}\\$\boldsymbol{({\cal P}_2)}$} & 
        \shortstack{\textbf{Axion-1}\\$\boldsymbol{(\chi_1)}$} & 
        \shortstack{\textbf{Axion-2}\\$\boldsymbol{(\chi_2)}$} & 
        \shortstack{\textbf{Illustr.}\\~} & 
        \shortstack{\textbf{Success}\\~}
        \\ \hline
        $\cal N_\text{fields}$ &
        1 &
        1 &
        2 &
        2 &
        3 &
        3 
        \\ \hline
        $V_0~[\times 10^{-9}]$ &
        $2.21$ &
        $1.89$ &
        $1.89$ &
        $1.89$ &
        $1.89$ &
        $2.07$
        \\
        $C_2$ & 
        $0.5$ &
        $0.5$ &
        $0.5$ &
        $0.5$ &
        $0.5$ &
        $0.5$
        \\
        $C_3$ & 
        $0.2661248$ &
        $0.26614$ &
        $0.26614$ &
        $0.26614$ &
        $0.26614$ &
        $0.26614$
        \\
        $C_4~[\times 10^{-4}]$ & 
        $7.82$ &
        $7.00$ &
        $7.00$ &
        $7.00$ &
        $7.00$ &\
        $7.00$
        \\
        $C_5$ & 
        $0.03911638$ &
        $0.0391168$ &
        $0.0391168$ &
        $0.0391168$ &
        $0.0391168$ &
        $0.0391168$
        \\
        $C_6$ & 
        $0.03575049$ &
        $0.0357523$ &
        $0.0357523$ &
        $0.0357523$ &
        $0.0357523$ &
        $0.0357523$
        \\ \hline
        $M_1~[\times 10^{-7}]$ & 
        $-$ &
        $-$ &
        $8$ &
        $-$ &
        $8$ &
        $8$
        \\
        $f_1$ & 
        $-$ &
        $-$ &
        $0.5$ &
        $-$ &
        $0.5$ &
        $0.1$
        \\
        $M_2~[\times 10^{-7}]$ & 
        $-$ &
        $-$ &
        $-$ &
        $5$ &
        $5$ &
        $8$
        \\
        $f_2$ & 
        $-$ &
        $-$ &
        $-$ &
        $0.5$ &
        $0.5$ &
        $0.5$
        \\ \hline
        $\varphi_i$ &
        $10$ &
        $8$ &
        $8$ &
        $8$ &
        $8$ &
        $8$
        \\
        $\chi_{1,i}$ &
        $-$ &
        $-$ &
        $0.1\pi f_1$ &
        $-$ &
        $0.1\pi f_1$ &
        $0.1\pi f_1$
        \\
        $\chi_{2,i}$ &
        $-$ &
        $-$ &
        $-$ &
        $0.1\pi f_2$ &
        $0.1\pi f_2$ &
        $0.1\pi f_2$
    \end{tabular}
    \caption{Parameter values for all models considered in this work, for both the inflaton potential of Eq.~\eqref{eq:V-inf-new} and the axion potential of Eq.~\eqref{eq:V-ax}. The parameter $C_1$ is always found by requiring $V_{\rm inf}=0$ at $\varphi=0$, such that $C_1 = 1 - C_2/(1-C_3) - C_4 + C_5/(1+C_6)$. All values are given in Planck units.
    All the axion models adopt the $\calP_2$ parameters for $V_{\rm inf}$: the ``Axion-1'', ``Axion-2'' and ``Illustrative'' models are used for illustrative purposes, while the ``Success'' model achieves consistency with CMB observables while also producing PBHs in the asteroid-mass range. 
    }
    \label{tab:all-params}
\end{table}

\begin{table}[ht!]
    \centering
    \begin{tabular}{@{\extracolsep{3pt}}c|*{3}c}
        \shortstack{\textbf{Model}\\~} & 
        \shortstack{\textbf{Reference}\\$\boldsymbol{({\cal P}_1)}$} & 
        \shortstack{\textbf{No-PBH}\\$\boldsymbol{({\cal P}_2)}$} & 
        \shortstack{\textbf{Success}\\~}
        \\ \hline
        $A_s~[\times 10^{-9}]$ &
        $2.10$ &
        $2.10$ &
        $2.10$ 
        \\
        $n_s$ &
        $0.9651$ &
        $0.9752$ &
        $0.9653$
        \\
        $\alpha_s$ & 
        $-0.003$ &
        $-0.003$ &
        $-0.003$
        \\
        $r$ & 
        $0.022$ &
        $0.001$ &
        $0.002$
        \\
        $\beta_{\rm iso}$ & 
        $-$ &
        $-$ &
        $0.0001$
        \\
        $f_{\rm NL}^{\rm equil}$ & 
        $109$ &
        $0.60$ &
        $0.91$
        \\
        $M_{\rm PBH}~[{\rm g}]$ & 
        $1.21\times 10^{22}$ &
        $-$ &
        $2.53\times 10^{21}$
        \\ \hline
        $\prk^{\rm peak}$ & 
        $7.17\times 10^{-3}$ &
        $1.87\times 10^{-4}$ &
        $3.50\times 10^{-3}$
    \end{tabular}
    \caption{Observables for the relevant models considered in this work, as defined in Tab.~\ref{tab:all-params}. CMB scales exit at $N_{\rm CMB}=53$, as discussed in Ref.~\cite{Cicoli:2022sih}. Observables are computed using \texttt{PyTransport}~\cite{Mulryne:2016mzv,Ronayne:2017qzn}.
    }
    \label{tab:all-obs}
\end{table}

\begin{figure}[ht!]
    \centering
    \includegraphics[width=\linewidth]{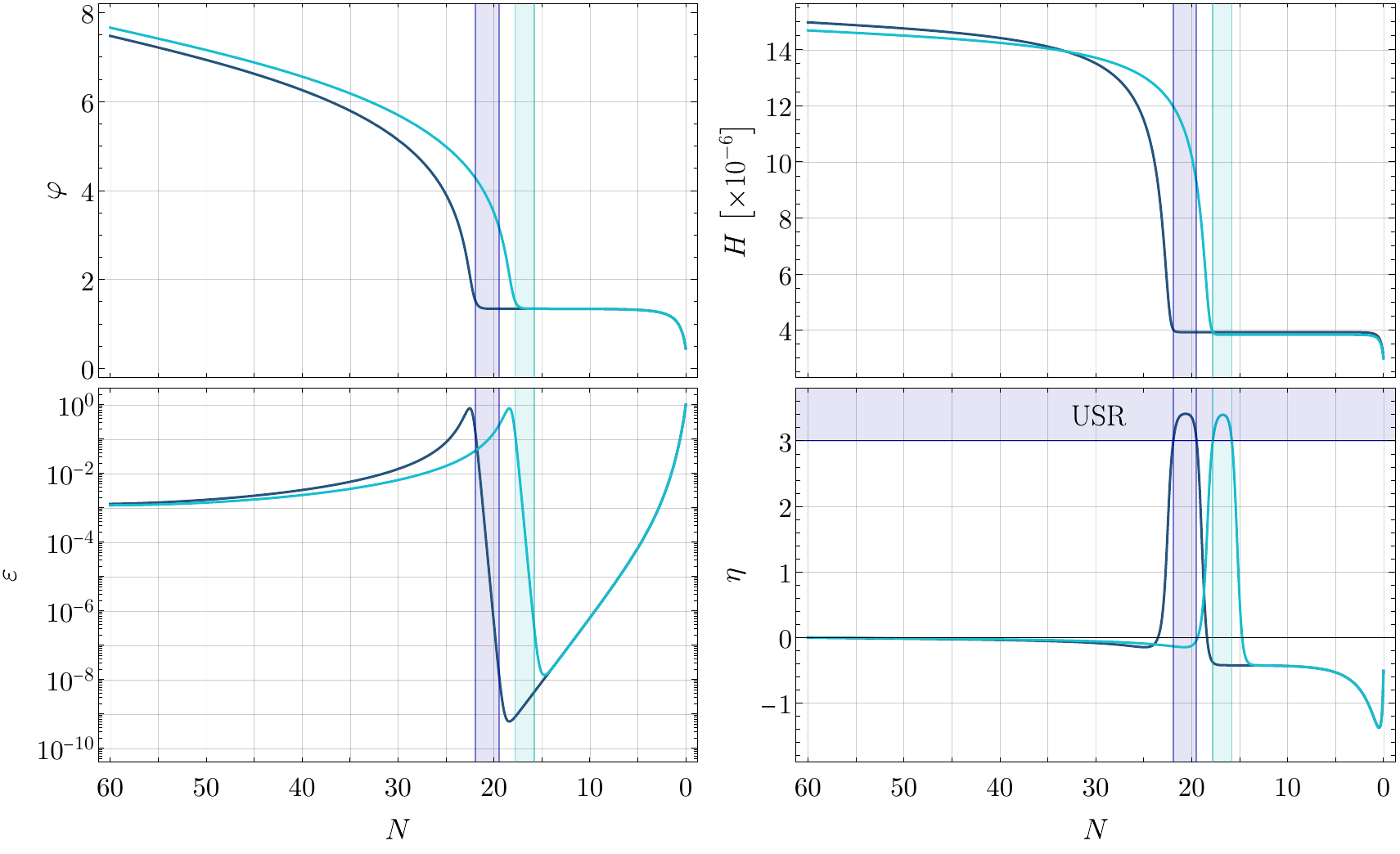}
    \caption{
    Evolution of the background quantities for the single-field Fibre Inflation model of Eq.~\eqref{eq:V-inf}, for the $\calP_1$ (blue) and the $\calP_2$ (cyan) inflaton parameters given in Tab.~\ref{tab:all-params}. It can be appreciated that the ``No-PBH'' model exhibits a shorter phase of ultra-slow-roll evolution ($\Delta N_{\rm USR}=2.0$ for $\calP_2$, in cyan shading, against $\Delta N_{\rm USR}=2.4$ for $\calP_1$, in blue shading), which translates to a smaller enhancement of the peak of the curvature power spectrum (see Fig.~\ref{fig:PRk-SF}).
    }
    \label{fig:SF-bckgr}
\end{figure}

\begin{figure}[ht!]
    \centering
    \includegraphics[width=\linewidth]{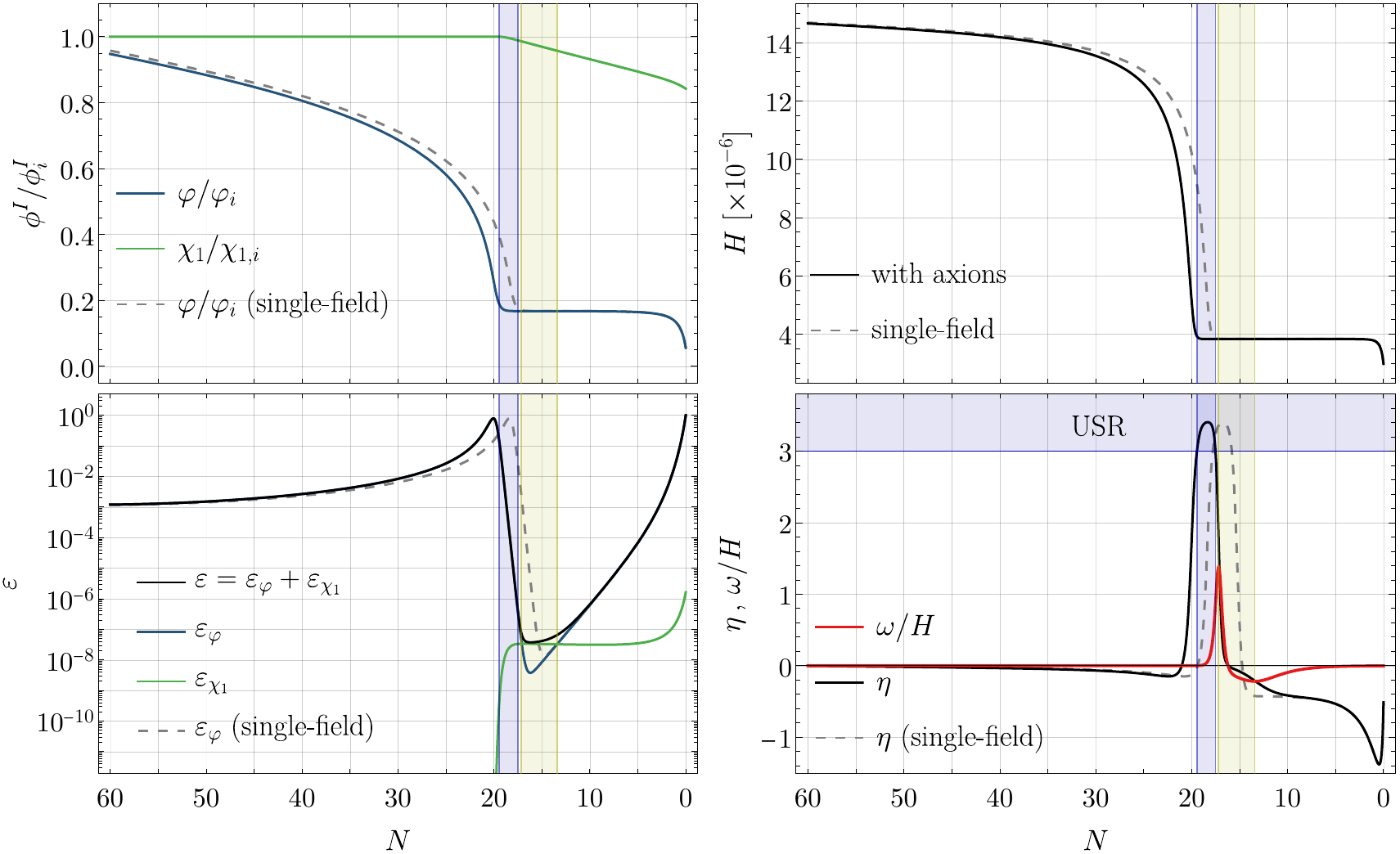}
    \caption{
    Evolution of the background quantities for the two-field Fibre Inflation model of Eq.~\eqref{eq:action-full} with axion $\chi_1$ only, for the ``Axion-1'' parameters given in Tab.~\ref{tab:all-params}.
    }
    \label{fig:ax1-bckgr}
\end{figure}

\begin{figure}[ht!]
    \centering
    \includegraphics[width=\linewidth]{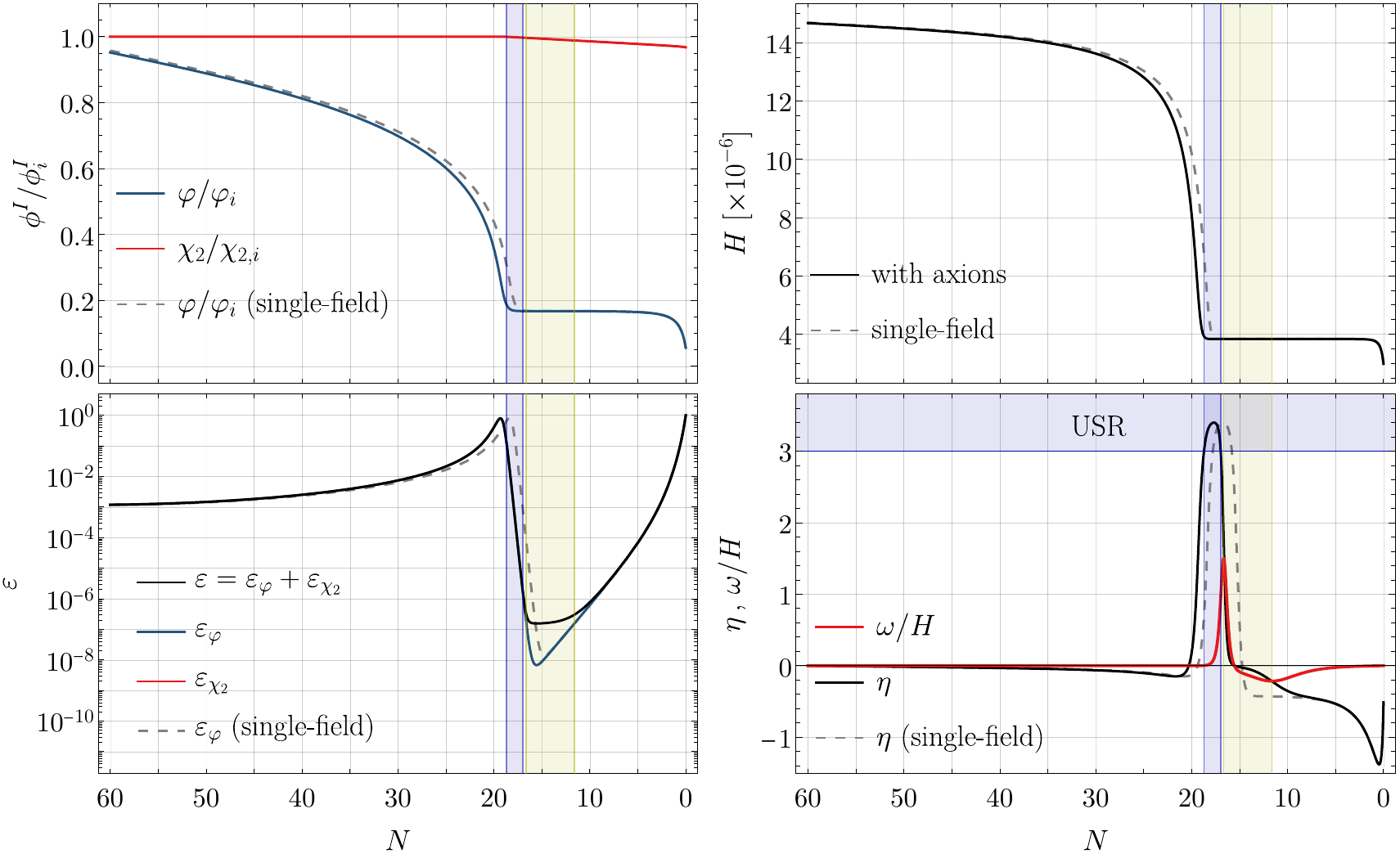}
    \caption{
    Evolution of the background quantities for the two-field Fibre Inflation model of Eq.~\eqref{eq:action-full} with axion $\chi_2$ only, for the ``Axion-2'' parameters given in Tab.~\ref{tab:all-params}.
    }
    \label{fig:ax2-bckgr}
\end{figure}

\begin{figure}[ht!]
    \centering
    \includegraphics[width=\linewidth]{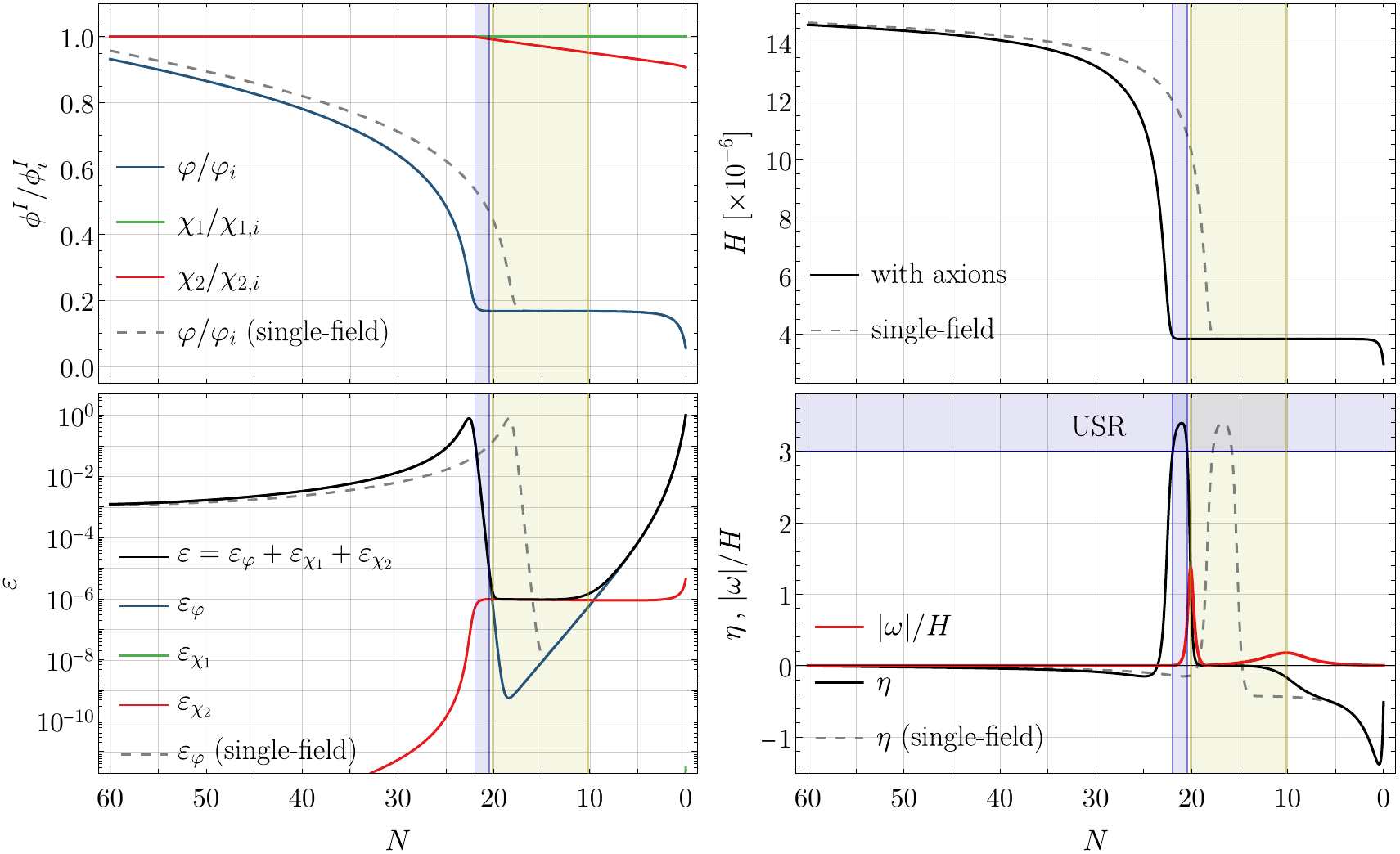}
    \caption{
    Evolution of the background quantities for the three-field Fibre Inflation model of Eq.~\eqref{eq:action-full}, for the ``Success'' parameters given in Tab.~\ref{tab:all-params}.
    }
    \label{fig:SUC-bckgr}
\end{figure}

The background evolution of the single-field models is shown in Fig.~\ref{fig:SF-bckgr}; that of the ``Axion-1'' and ``Axion-2'' models is shown in Figs.~\ref{fig:ax1-bckgr} and~\ref{fig:ax2-bckgr}; that of the ``Success'' model is shown in Fig.~\ref{fig:SUC-bckgr}.

\bibliographystyle{JHEP}
\bibliography{refs}

\end{document}